\documentclass[aps,prd,
floatfix,nofootinbib,groupedaddress,showpacs,longbibliography]{revtex4-2}

\usepackage{amsmath}
\usepackage{amssymb}
\usepackage{graphicx}

\usepackage{latexsym}
\usepackage{array}
\usepackage{verbatim}

\usepackage{rotating}
\usepackage{color}

\usepackage[utf8]{inputenc}
\usepackage{ulem}

\def\lsim{\mathrel{\raise.3ex\hbox{$<$\kern-.75em\lower1ex\hbox{$\sim$}}}}
\def\gsim{\mathrel{\raise.3ex\hbox{$>$\kern-.75em\lower1ex\hbox{$\sim$}}}}

\newcommand{\be}{\begin{equation}}
\newcommand{\ee}{\end{equation}}

\newlength{\absize}
\def\lsim{\mathrel{\rlap{\raise 2.5pt \hbox{$<$}}\lower 2.5pt
\hbox{$\sim$}}}

\allowdisplaybreaks

\begin{document}

\title{Bounds on the mass and mixing of $Z^\prime$ and $W^\prime$ bosons decaying into different pairings of $W$, $Z$, or Higgs bosons using CMS data at the LHC}

\author{P. Osland}
\email{Per.Osland@uib.no}
\affiliation{Department of Physics and
Technology, University of Bergen, Postboks 7803, N-5020 Bergen,
Norway}
\author{A.~A. Pankov}
\email{pankov@ictp.it}
\affiliation{The Abdus Salam ICTP
Affiliated Centre, Technical University of Gomel, 246746 Gomel,
Belarus}
\affiliation{Institute for Nuclear Problems, Belarusian
 State University, 220030 Minsk, Belarus}
 \affiliation{Joint Institute  for Nuclear Research, Dubna 141980 Russia}
\author{I.~A. Serenkova}
\email{Inna.Serenkova@gmail.com}
\affiliation{The Abdus Salam ICTP Affiliated
Centre, Technical University of Gomel, 246746 Gomel, Belarus}

\date{\today}

\begin{abstract}
The full CMS Run~2 datasets with time-integrated luminosity of 137 fb$^{-1}$ in
the diboson channels are used to probe benchmark models with extended gauge sectors
such as $E_6$, left-right symmetric (LR) and the sequential
standard model (extended gauge model, EGM),
that predict the existence of neutral $Z'$- and charged $W'$-bosons decaying to a pair of bosons $WW$, $ZH$, $WZ$ and $WH$ in the semileptonic final state. These benchmark models are used to
interpret the results. Exclusion limits at the 95\% C.L. on the
$Z'$ and $W'$ resonance production cross section times branching
ratio to electroweak gauge boson pairs in
the resonance  mass range between 1.0 and 4.5  TeV  are here converted to
constraints on $Z$-$Z'$ and $W$-$W'$ mixing parameters and
masses. We present exclusion regions on the
parameter spaces of the $Z'$ and $W'$ 
and show that the obtained exclusion regions are significantly extended compared to
those derived from the previous analysis performed with Tevatron
data as well as with the CMS data collected at 7 and 8 TeV in Run~1. The reported limits are the most restrictive to date.
\end{abstract}

\maketitle

\section{Introduction} \label{sec:I}
A variety of theoretical extensions to the standard model of particle physics (SM)
predict new phenomena in high-energy proton-proton ($pp$) collisions, the discovery of which is one of the main goals of the CERN Large Hadron Collider (LHC).
The LHC allows to probe new phenomena, new particles and interactions, at energies of several TeV. A wide range of models predicts the production of new heavy, TeV-scale, resonances or vector bosons decaying to pairs of SM electroweak vector bosons  (jointly referred to as $V$ in the following, with $V=W,\, Z$), and SM Higgs ($H$) bosons.
Models studied in the literature include extended
gauge models (EGM) \cite{Altarelli:1989ff,Eichten:1984eu,Hewett:1988xc,Leike:1998wr,Langacker:2008yv,Erler:2009jh}, models of warped extra dimensions \cite{Randall:1999ee,Davoudiasl:2000wi}, technicolour
models \cite{Lane:2002sm,Eichten:2007sx} associated with
technirho and other technimesons, composite
Higgs models \cite{Agashe:2004rs, Giudice:2007fh}, and the heavy
vector-triplet (HVT) model \cite{Pappadopulo:2014qza}, which generalises a large number of models that predict spin-1 neutral ($Z'$) and charged ($W'$) resonances.

The extended gauge models are among the best motivated theoretical
scenarios beyond the SM that predict the existence of new heavy neutral
and charged vector bosons ($Z^\prime$ and $W^\prime$)
\cite{ParticleDataGroup:2020ssz,Langacker:2008yv}.
These models are considered as benchmark scenarios for diboson resonances having spin~1
($W' \to WZ$ or $WH$, $Z' \to WW$ or $ZH$), produced predominantly via quark-antiquark annihilation ($q{\bar{q}^\prime} \to W'$, $q\bar q \to Z'$). 

The neutral and charged massive resonance production at hadron level  and its subsequent decay to pairs of electroweak gauge and Higgs bosons can be expressed as
\begin{subequations}
\begin{eqnarray}
pp&\to &Z^\prime X\to WW\, X  \label{ww}\;, \\
pp&\to &Z^\prime X\to ZH\, X  \label{zh}\;,
\end{eqnarray}
\end{subequations}
and
\begin{subequations}
\begin{eqnarray}
pp&\to &W^\prime X\to WZ\, X  \label{wz}\;, \\
pp&\to &W^\prime X\to WH\, X  \label{wh}\;.
\end{eqnarray}
\end{subequations}
Depending on the mass and the couplings to the SM  quarks and final electroweak bosons, these new states could be accessible to the LHC and observable by the ATLAS and CMS experiments \cite{Salvioni:2009mt,Gulov:2018zij}.

In the simplest models under study such as the Sequential Standard Model (SSM) \cite{Altarelli:1989ff} new neutral $Z'_{\rm SSM }$ and charged $W'_{\rm SSM }$ bosons have couplings to fermions that are identical to those of the SM $Z$ and $W$ bosons, but for which the trilinear couplings $Z'WW$ and $W'WZ$ are absent, $g_{Z'WW}=0$ and $g_{W'WZ}=0$.
This suppression may arise naturally in an EGM: if the new gauge bosons and the SM ones belong to different gauge groups, a vertex such as $Z'WW$ $(W'WZ)$ is forbidden. They can only be induced after symmetry breaking due to mixing of the gauge eigenstates.

 Another class of  models considered here are those inspired by Grand Unified Theories (GUT), which are motivated by gauge unification or a restoration of the left--right symmetry violated by the weak interaction. Examples considered in this paper include the $Z^\prime$ bosons of the $E_6$-motivated~\cite{Langacker:2008yv} theories containing
$Z^\prime_\psi$, $Z^\prime_\eta$, $Z^\prime_\chi$; and high-mass neutral bosons of the left-right (LR) symmetric extensions of the SM, based on the $SU(2)_L\bigotimes SU(2)_R\bigotimes U(1)_{B-L}$ gauge group, where $B-L$ refers to the difference between baryon and lepton numbers.

The properties of possible $Z'$ and $W'$ bosons are also constrained by measurements
of electroweak (EW) \cite{Erler:2009jh}  processes at low energies, i.e., at energies much below their masses. Such bounds on the $Z$-$Z'$ ($W$-$W'$) mixing are mostly due to the
constraints on deviation in $Z$ ($W$) properties from the SM predictions.
In particular,  limits from direct hadron production with subsequent diboson decay at the Tevatron \cite{Aaltonen:2010ws} and from  virtual
effects at LEP, through interference or mixing with the $Z$ boson,
imply that any new $Z^{\prime}$ boson is rather heavy and mixes
very little with the $Z$ boson.  The
measurements show that the mixing angles, referred to as  $\xi_{Z\text{-}Z^\prime}$ and $\xi_{W\text{-}W^\prime}$,  between the gauge eigenstates must be smaller than about $10^{-3}$ and $10^{-2}$, respectively \cite{ParticleDataGroup:2020ssz,Erler:2009jh}.

Previous analyses of the $Z$-$Z'$ and $W$-$W'$ mixing \cite{Osland:2017ema,Bobovnikov:2018fwt,Serenkova:2019zav}
were carried out using the diboson  production data set corresponding to the time-integrated luminosity of $\sim$ 36 fb$^{-1}$  collected in 2015 and 2016 with the ATLAS and CMS  collaborations at $\sqrt{s}=$ 13~TeV where, in the former case, electroweak $Z$ and $W$ gauge bosons decay into the semileptonic channel \cite{Aaboud:2017fgj} or into the dijet final state \cite{Sirunyan:2017acf}. Further updated results were obtained
using the diboson and dilepton Run~2 production data set
corresponding to an integrated luminosity of 139 fb$^{-1}$ \cite{Pankov:2019yzr,Osland:2020onj} recorded  by the ATLAS detector  \cite{Aad:2020ddw}. In the analysis presented here, we utilize the full Run~2 CMS data set on diboson resonance production published recently in Refs.~\cite{CMS:2021klu,hepdata102645,CMS:2021fyk,hepdata101374} for the $VV$ and $VH$ channels corresponding to an integrated luminosity of 137 fb$^{-1}$. The present analysis includes various $Z'$ models such as  EGM (SSM), $E_6$ based $Z_\chi$, $Z_\psi$, $Z_\eta$, as well as the $Z_{\rm LR}$ boson appearing in models with left-right symmetry. Also, a new set of diboson processes, $W'\to WH$ and $Z'\to ZH$, were examined.\footnote{We do not consider here effects of bosonic mixing in dilepton production \cite{Osland:2020onj} as it is out of scope
of the present paper.} Thus, our present analysis is complementary to the previous studies performed for ATLAS data in \cite{Osland:2020onj}.

 We present results as constraints on the relevant $Z$-$Z'$
($W$-$W'$) mixing angle, $\xi_{Z\text{-}Z^\prime}$ ($\xi_{W\text{-}W^\prime}$),  and on the mass $M_{Z'}$ $(M_{W'})$ and display the
combined allowed parameter space for the benchmark $Z'$ ($W'$) boson models,
showing also indirect constraints from electroweak precision data (EW), previous direct search constraints from the Tevatron and from the LHC with 7 and 8~TeV in Run~1 (where available), as well as those obtained from the LHC at 13 TeV with the full CMS Run~2 data set of time-integrated luminosity of 137 fb$^{-1}$ in the semileptonic \cite{CMS:2021klu,hepdata102645,CMS:2021fyk,hepdata101374} final states.

The paper is organized as follows. In Sect.~\ref{sect:mixing} we present the theoretical framework, then, in Sects.~\ref{sect:productionWp} and \ref{sect:productionZp} we review the production and decay of $W^\prime$ and $Z^\prime$, respectively. Finally, Sect.~\ref{sect:conclusions} contains concluding remarks. The paper is a follow-up study of our earlier analysis of the corresponding ATLAS data \cite{Osland:2020onj}. In order to make it self-contained, there is some repetition of basic formulas.

\section{$V$--$V^\prime$ mixing } 
\label{sect:mixing}

As mentioned above, in the SSM, the coupling constants of the
$W'$ and $Z'$ bosons with SM fermions are identical to the
corresponding SM couplings, while the $W'$ and $Z'$ couplings to, respectively,  $WZ$ and $WW$
vanish, $g_{W'WZ}=g_{Z'WW}=0$. Such a suppression may arise in an EGM in a natural
manner: if the new gauge bosons and those of the SM belong to
different gauge groups, vertices such as $W'WZ$ and $Z'WW$ do not arise. They
can only occur after symmetry breaking due to mixing of the gauge
eigenstates. Triple gauge boson couplings (such as $W'WZ$ and $Z'WW$) as well
as the vector-vector-scalar couplings (like $W'WH$ and $Z'ZH$) arise from the
symmetry breaking and may contribute to the $W'$ and $Z'$ decays, respectively.
The vertices are then suppressed by a factor of the order of
$(M_W/M_{V'})^2$, where $V^{\prime}$  represents  a $W^{\prime}$ or a $Z^{\prime}$ boson.

In an EGM \cite{Altarelli:1989ff}, the trilinear gauge boson couplings are modified by mixing
factors
\begin{equation} \label{Eq:define-xi}
\xi_{V\text{-}V'}={\cal C} \times (M_W/M_{V'})^2,
\end{equation}
where ${\cal C}$ is a scaling constant that sets the coupling
strength. Note that the EGM can be parametrized either in terms of $(M_{V'},{\cal C})$ or in terms of $(M_{V'},\xi_{V-V'})$. Specifically,
in an EGM the standard-model trilinear gauge boson coupling
strength $g_{WWZ}$ ({$=e\cot\theta_W$}), is replaced by
$g_{W'WZ}=\xi_{W\text{-}W^\prime}\cdot g_{WWZ}$ in the $WZ$ channel and  $g_{Z'WW}=\xi_{Z\text{-}Z^\prime}\cdot g_{WWZ}$  in the $WW$ channel.
Following the parametrization of the trilinear gauge boson couplings $W'WZ$ and $Z'WW$ presented in \cite{Aaltonen:2010ws}
for the analysis and interpretation of the
CDF data on  $p\bar{p}\to W'X\to WZX$  and $p\bar{p}\to Z'X\to W^+W^-X$, expressed in terms of
two free parameters,\footnote{Such $W'$ and $Z'$, described in terms of
two parameters (mass and mixing), are here referred to as EGM bosons.}
$\xi_{W\text{-}W^\prime}$ ($\xi_{Z\text{-}Z^\prime}$) and $M_{W'}$ ($M_{Z'}$), we will set two-dimensional limits,  by using
the CMS resonant diboson production data \cite{CMS:2021klu,hepdata102645,CMS:2021fyk,hepdata101374}
collected in the full Run~2 data set with time-integrated luminosity of  137 fb$^{-1}$.
The presented analysis in the EGM with two free parameters is more
general than the previous ones where the only parameter is the
$V'$ mass. As for the SSM, one has $V'_{\rm SSM}\equiv V'_{\rm EGM}$ when $\xi_{V\text{-}V'}=0$.

Note that the parametrization of boson mixing introduced by Altarelli et al. \cite{Altarelli:1989ff}, though being simplified, has a well-motivated theoretical basis. To be specific,  we briefly consider $Z^0$--$Z^{0\prime}$ mixing within the framework of  models with extended gauge sector (see, e.g.
\cite{Hewett:1988xc,Leike:1998wr,Langacker:2008yv,Erler:2009jh}).
The mass eigenstates $Z$ and $Z^\prime$ are admixtures of the weak eigenstates $Z^0$ of $SU(2)\times U(1)$ and $Z^{0\prime}$ of the extra $U(1)'$, respectively:
\begin{subequations}
\label{Eq:Z12-couplings}
\begin{eqnarray}
&Z& = Z^0\cos\phi + Z^{0\prime}\sin\phi\;, \label{z} \\
&Z^\prime& = -Z^0\sin\phi + Z^{0\prime}\cos\phi\;. \label{zprime}
\end{eqnarray}
\end{subequations}
In each case there is a relation between the $Z^0$-$Z^{0\prime}$ mixing angle $\phi$ and the masses $M_Z$ and $M_{Z'}$ \cite{Langacker:2008yv}:
\begin{equation} \label{phi}
\tan^2\phi={\frac{M_{Z^0}^2-M_Z^2}{M_{Z'}^2-M_{Z^0}^2}}\simeq
\frac{2\, M_{Z^0}\,\Delta M_{Z^0Z}}{M_{Z'}^2}\;,
\end{equation}
where the downward shift $\Delta M_{Z^0Z}=M_{Z^0}-M_Z>0$, and
${M_{Z^0}}$ is the mass of the $Z$ boson in the absence of mixing, i.e., for
$\phi=0$, given by
\begin{equation} \label{MZ0}
{M_{Z^0}}=\frac{M_W}{\sqrt{\rho_0}\cos\theta_W},
\end{equation}
in terms of the charged ($M_W$) gauge boson mass and the $\rho_0$ parameter.
The mixing angle $\phi$ will play an important role in our analysis. Such mixing effects reflect the underlying gauge symmetry and/or the structure of the Higgs sector of the model as the
$\rho_0$ parameter depends on the ratios of the Higgs vacuum expectation
values and on the total and third components of weak isospin of the Higgs fields.
For each type of $Z^{0\,\prime}$ boson, defined by its gauge couplings, there are three classes of models, which differ in the assumptions concerning the quantum numbers of the Higgs fields which generate the $Z$-boson mass matrix \cite{Hewett:1988xc,Leike:1998wr,Langacker:2008yv}:

\begin{itemize}
\item[(i)]
The least constrained ($\rho_0$ free) model makes no assumption concerning the
Higgs sector. It allows arbitrary $SU(2)$ representations
for the Higgs fields, and is the analogue of allowing  $\rho_0\neq 1$  in
the $SU(2)\times U(1)$ model. In this case $M_Z$, $M_{Z'}$ and $\phi$ are all free parameters.

\item[(ii)]
If one assumes that all $SU(2)$ breaking is due to Higgs doublets and singlets  ($\rho_0=1$ model), there are only two free parameters, which we identify as $\phi$ and $M_{Z^\prime}$. We will adopt this parametrization throughout the paper.

\item[(iii)]
Finally, in specific models one specifies not only the $SU(2)$ assignments
but the $U(1)^\prime$ assignments of the Higgs fields.
Since the same Higgs multiplets generate both $M_Z$ and $\phi$, one
has an additional constraint.    To a good approximation,
for $M_Z\ll M_{Z'}$, in specific ``minimal-Higgs models'', one has an
additional constraint \cite{Langacker:1991pg,Djouadi:1991sx}
\begin{equation}\label{phi0}
\phi\simeq -s^2_\mathrm{W}\
\frac{\sum_{i}\langle\Phi_i\rangle{}^2I^i_{3L}Q^{\prime}_i}
{\sum_{i}\langle\Phi_i\rangle^2(I^i_{3L})^2} =
P\,{\frac{\displaystyle M^2_Z}{\displaystyle M^2_{Z'}}},
\end{equation}
where $s_\mathrm{W}$ is the sine of the electroweak  angle.
In these models $\phi$ and $M_{Z'}$ are not independent and
there is only one (e.g., $M_{Z'}$) free parameter. This parametrization is of the form presented in Eq.~(\ref{Eq:define-xi}).
Furthermore, $\langle\Phi_i\rangle$ are the Higgs (doublet) vacuum expectation values
spontaneously breaking the symmetry, and $Q^\prime_i$  are their
charges with respect to the additional $U(1)'$. In
these models the same Higgs multiplets are responsible for both
generation of the mass $M_Z$ and for the strength of the
$Z^0$-$Z^{0\prime}$ mixing. Thus $P$ is a model-dependent
constant.
\end{itemize}

This $Z^0$-$Z^{0\prime}$ mixing induces a change in the couplings of the two bosons to fermions. From Eq.~(\ref{Eq:Z12-couplings}), one obtains the vector and
axial-vector couplings of the $Z$ and $Z'$ bosons to fermions:
\begin{subequations} \label{v2}
\begin{alignat}{2}
v_{f} &= v^0_f\cos\phi + v^{0\prime}_f\sin\phi\;, &\quad
a_{f} &= a^0_f \cos\phi + a^{0\prime}_f \sin\phi\;, \label{v1} \\
v^\prime_{f} &= v_f^{0\prime} \cos\phi - v^0_f \sin\phi\;, &\quad
a^\prime_{f} &= a_f^{0\prime} \cos\phi -a^0_f \sin\phi\;,
\end{alignat}
\end{subequations}
with unprimed and primed couplings referring to $Z^{0}$
and $Z^{0\prime}$, respectively, and found, e.g. in~\cite{Hewett:1988xc,Leike:1998wr,Langacker:2008yv}.

An important property of the models
under consideration is that the gauge eigenstate $Z^{0\prime}$ does
not couple to the $W^+W^-$ pair since it is neutral under
$SU(2)$. Therefore the $W$-pair production is sensitive to a
$Z^\prime$ only to the extent that there is a non-zero $Z^0$-$Z^{0\prime}$ mixing.
From Eq.~(\ref{Eq:Z12-couplings}), one obtains:
\begin{subequations}
\begin{eqnarray}
&& g_{WWZ}=\cos\phi\;g_{WWZ^{0}}\;, \label{WWZ1} \\
&& g_{WWZ'}=-\sin\phi\; g_{WWZ^{0}}\;,\label{WWZ2}
\end{eqnarray}
\end{subequations}
where $g_{WWZ^{0}}=e\cot\theta_W$. Also, $g_{WW\gamma}=e$.

In many extended gauge models, while the couplings to fermions are
not much different from those of the SM, the $Z'WW$ coupling is
substantially suppressed with respect to that of the SM. In fact,
in the extended gauge models  the SM trilinear gauge boson
coupling strength, $g_{WWZ^{0}}$, is replaced by $g_{WWZ^{0}} \rightarrow
\xi_{Z-Z'}\cdot g_{WWZ^{0}}$, where $\xi_{Z-Z'} \equiv \vert\sin\phi\vert$
(see Eq.~(\ref{WWZ2})) is the mixing factor.
We will set cross section limits on such $Z'$ as functions of the mass $M_{Z'}$ and $\xi_{Z-Z'}$.

In addition, we study $W$-$W^{\prime}$ mixing in the processes 
 (\ref{wz}) and (\ref{wh})
within the framework of the EGM model \cite{Altarelli:1989ff,Eichten:1984eu}.
Mass mixing may be induced between the electrically charged gauge bosons at the tree level. The physical (mass) eigenstates of $W$ and $W^\prime$ are admixtures of the weak eigenstates denoted as $\hat{W}$ and $\hat{W'}$, respectively, and obtained  by a rotation of those fields \cite{ParticleDataGroup:2020ssz,Grojean:2011vu,Dobrescu:2015yba}:
\begin{subequations}
\begin{eqnarray}
W^\pm & = & \hat{W}^\pm \cos\theta_{WW'} + \hat{W'}^\pm \sin\theta_{WW'},  \\
W'^\pm & = & -\hat{W}^\pm \sin\theta_{WW'} + \hat{W'}^\pm \cos\theta_{WW'},
\end{eqnarray}
\end{subequations}
in analogy with Eq.~(\ref{Eq:Z12-couplings}).
Upon diagonalization  of their mass matrix, the couplings of the observed $W$ boson are shifted from the SM values. The mixing parameter $\xi_{W-W'}$ between gauge eigenstates can be defined as $\xi_{W-W'} \equiv \vert\sin\theta_{WW'}\vert$.

\section{Hadron production and decay of $W'$ boson}
\label{sect:productionWp}

In this section, we consider the simplest EGM model which predicts charged heavy gauge bosons.
At the lowest order in the EGM, $W'$ production and decay into $WZ$ and $WH$ in
proton-proton collisions occur through quark-antiquark
annihilation in the $s$-channel.
Adopting the Narrow-Width Approximation (NWA), one can factorize the
processes (\ref{wz}) and (\ref{wh}) into the $W'$ production and the $W'$
decay,
\begin{subequations}
\begin{align}
\sigma(pp\to W' X\to WZ\,X)&  = \sigma(pp\to W'\,X) \times \text{BR}(W' \to
WZ)\;\label{sigwz}\;, \\
\sigma(pp\to W' X\to WH\,X)&  = \sigma(pp\to W'\,X) \times \text{BR}(W' \to
WH)\;\label{sigwh}\;.
\end{align}
\end{subequations}
Here, $\sigma(pp\to W' X)$ is the  total (theoretical) $W'$
production cross section,
$\text{BR}(W' \to WZ)=\Gamma_{W'}^{WZ}/\Gamma_{W'}$
and similarly
$\text{BR}(W' \to WH)=\Gamma_{W'}^{WH}/\Gamma_{W'}$
with
$\Gamma_{W'}$ the total width of the $W'$.

\subsection{The $W'$ width}
\label{sect:Wprime-width}

In the EGM the $W'$ bosons can decay into pairs of SM fermions (charged leptons, neutrinos and quarks), gauge bosons $WZ$ and $WH$. Specifically, in the calculation
of the total width $\Gamma_{W'}$ we consider the following
channels: $W'\to f{\bar{f}}^\prime$, $WZ$, and $WH$, where
$f$ is a SM fermion ($f=\ell,\nu,q$). Note, that here the $\ell$ includes $\tau$ leptons as well.
Only the familiar left-handed neutrinos are considered,
possible right-handed exotic neutrinos are assumed to be kinematically
unavailable as final states. Also,  we shall
ignore the couplings to other beyond-SM particles such as
SUSY partners and exotic fermions.
 As a result, the total decay
width of the $W^\prime$ boson is taken to be
\begin{equation}\label{gamma}
\Gamma_{W'} = \sum_f \Gamma_{W'}^{f\bar{f}'} + \Gamma_{W'}^{WZ} +
\Gamma_{W'}^{WH}.
\end{equation}

\begin{figure}[htb]
\begin{center}
\includegraphics[scale=0.5]{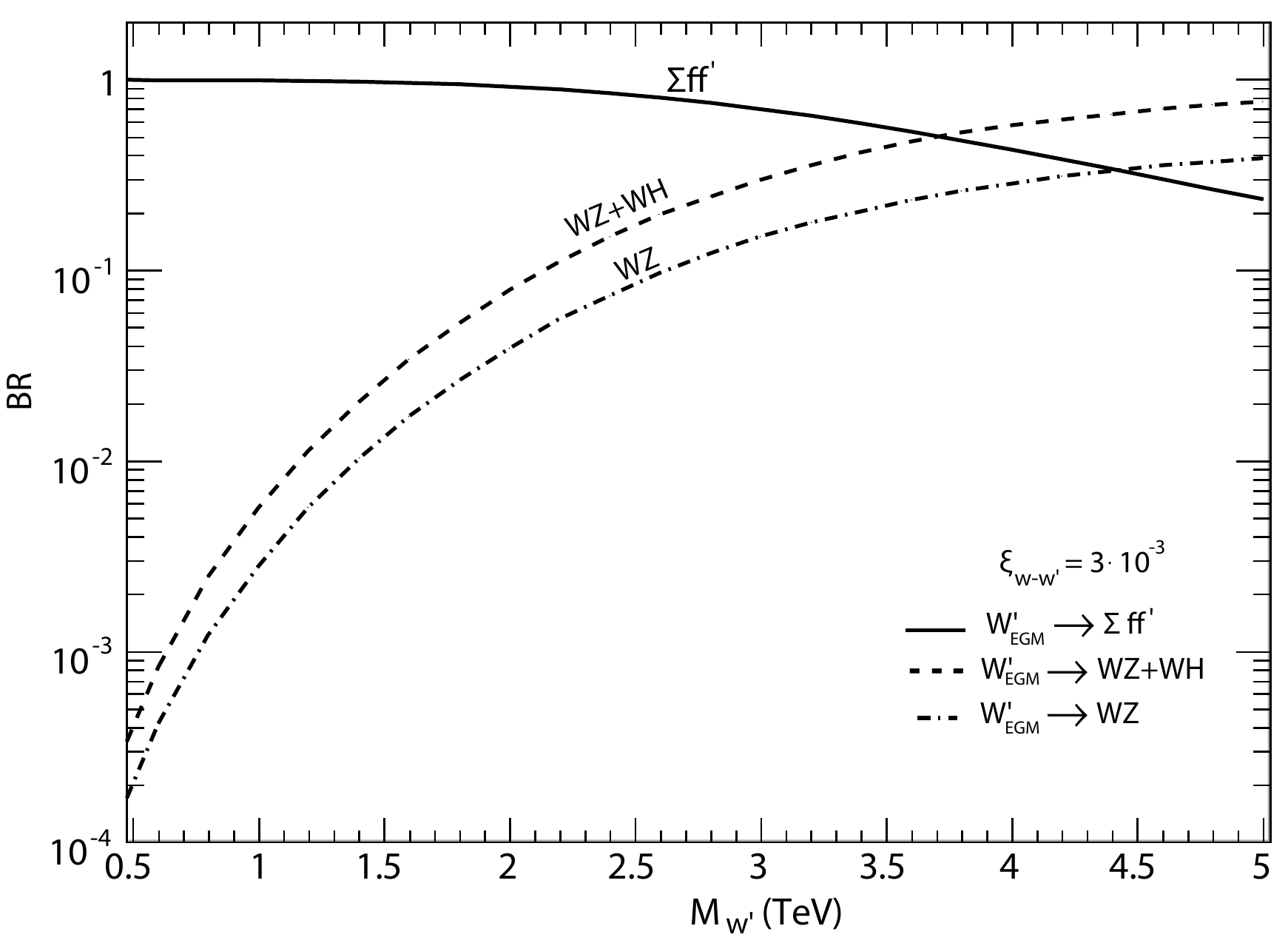}
\end{center}
\caption{
Branching ratios $\text{BR}(W'\to\sum{f\bar{f}'})$ (solid), $\text{BR}(W'\to WZ)$ (dash-dotted), and
$\text{BR}(W'\to WZ+WH)$ (dashed) vs $M_{W'}$ in the EGM where the $W$-$W'$ mixing factor is taken to be
$\xi_{W\text{-}W^\prime}=3\cdot 10^{-3}$.
It is assumed that $\text{BR}(W'\to WZ)=\text{BR}(W'\to WH)$.
}\label{fig:br-Wprime}
\end{figure}

The presence of the last two decay channels,  which are often
neglected at low and moderate values of $M_{W'}$, is due to
$W$-$W'$ mixing which is constrained to be tiny. In particular, for the range of $M_{W'}$ values below $\sim 2$ TeV,  the dependence
of $\Gamma_{W'}$ on the values of $\xi_{W\text{-}W^\prime}$ (within its allowed range)
induced by $\Gamma_{W'}^{WZ}$ and $\Gamma_{W'}^{WH}$ is
unimportant because $\sum_f \Gamma_{W'}^{f\bar {f'}}$ highly dominates
over the diboson partial widths as illustrated in Fig.~\ref{fig:br-Wprime} for a representative value of the mixing parameter. Therefore, in this mass range, one
can approximate the total width as $\Gamma_{W'} \approx \sum_f
\Gamma_{W'}^{f\bar {f'}}=3.5\%\times M_{W'}$ \cite{Serenkova:2019zav},
 where the sum runs over SM fermions only.
For heavier $W'$ bosons, the diboson decay channels, $WZ$ and
$WH$, start to play an important role,
and we are no longer able to ignore them \cite{Serenkova:2019zav,Pankov:2019yzr}.
To be specific,  we assume that both partial widths are comparable, $\Gamma_{W'}^{WH}\simeq
\Gamma_{W'}^{WZ}$ for heavy $M_{W'}$, as required by the
Equivalence theorem \cite{Chanowitz:1985hj}.

The partial width of the $W'\to WZ$ decay
channel in the EGM can be written as \cite{Altarelli:1989ff,Serenkova:2019zav,Osland:2020onj}:
\begin{eqnarray}
\label{GammaWZ}
 \Gamma_{W'}^{WZ}&=&\frac{\alpha_{\rm
em}}{48}\cot^2\theta_W\, M_{W'}
\frac{M_{W'}^4}{M_W^2M_Z^2}\left[\left(1-\frac{M_Z^2-M_W^2}{M_{W'}^2}\right)^2
-4\,\frac{M_W^2}{M_{W'}^2}\right]^{3/2} \\ \nonumber
&& \times\left[
1+10 \left(\frac{M_W^2+M_Z^2}{M_{W'}^2}\right) +
\frac{M_W^4+M_Z^4+10M_W^2M_Z^2}{M_{W'}^4}\right]\cdot\xi_{W\text{-}W'}^2.
\end{eqnarray}
For a fixed mixing factor $\xi_{W\text{-}W^\prime}$  and at large $M_{W'}$,
the total width
increases rapidly with the $W'$ mass because of the quintic
dependence  of the $WZ$ mode on the $W'$ mass
$\Gamma_{W'}^{WZ}\propto M_{W'}\left[{M_{W'}^4}/({M_W^2M_Z^2})\right]$,
 corresponding to the production of longitudinally polarized $W$ and $Z$ in the channel $W'\to
W_LZ_L$ \cite{Altarelli:1989ff}.
In this case, the $WZ$ mode (as well as $WH$) becomes dominant and
$\text{BR}(W' \to WZ)\to 0.5$, while the fermionic decay channels,
$\sum_f \Gamma_{W'}^{f\bar {f'}}\propto M_{W'}$, are increasingly
suppressed, as illustrated in Fig.~\ref{fig:br-Wprime}.

\subsection{Constraints on $W$-$W'$ mixing and  $M_{W^\prime}$}
\label{sect:Wprime-lim}

The data we consider were collected with the CMS detector during the 2015--2018 running period of the LHC, referred to as Run 2 and correspond to a time-integrated luminosity of 137~fb$^{-1}$. The CMS  experiment has presented the recent search for diboson resonances  based on the full Run 2 data set  in the semileptonic final states \cite{CMS:2021klu,hepdata102645,CMS:2021fyk,hepdata101374} and set limits on the $W'$
production cross sections times branching fraction in the processes
$pp\to W^\prime X\to WZ\, X$ and $pp\to W^\prime X\to WH\, X$ for $M_{W'}$ in the 1.0~TeV -- 4.5~TeV  range.

Such a search has also been presented by the ATLAS Collaboration  using  139~fb$^{-1}$ of data recorded for the $W^\prime \to WZ$ channel at $\sqrt{s}=13$ TeV  \cite{Aad:2020ddw}, and
exclusion limits at the 95\% C.L. on the
$W'$ resonance production cross section times branching
ratio to electroweak gauge boson pairs $WZ$ in
the resonance  mass range between $\simeq$ 0.5 TeV and 5 TeV. These data were similarly converted to
constraints on the $W$-$W'$ mixing parameter and $M_{W'}$ mass \cite{Osland:2020onj}.

\begin{figure}[hbt]
\begin{center}
\includegraphics[scale=0.52]{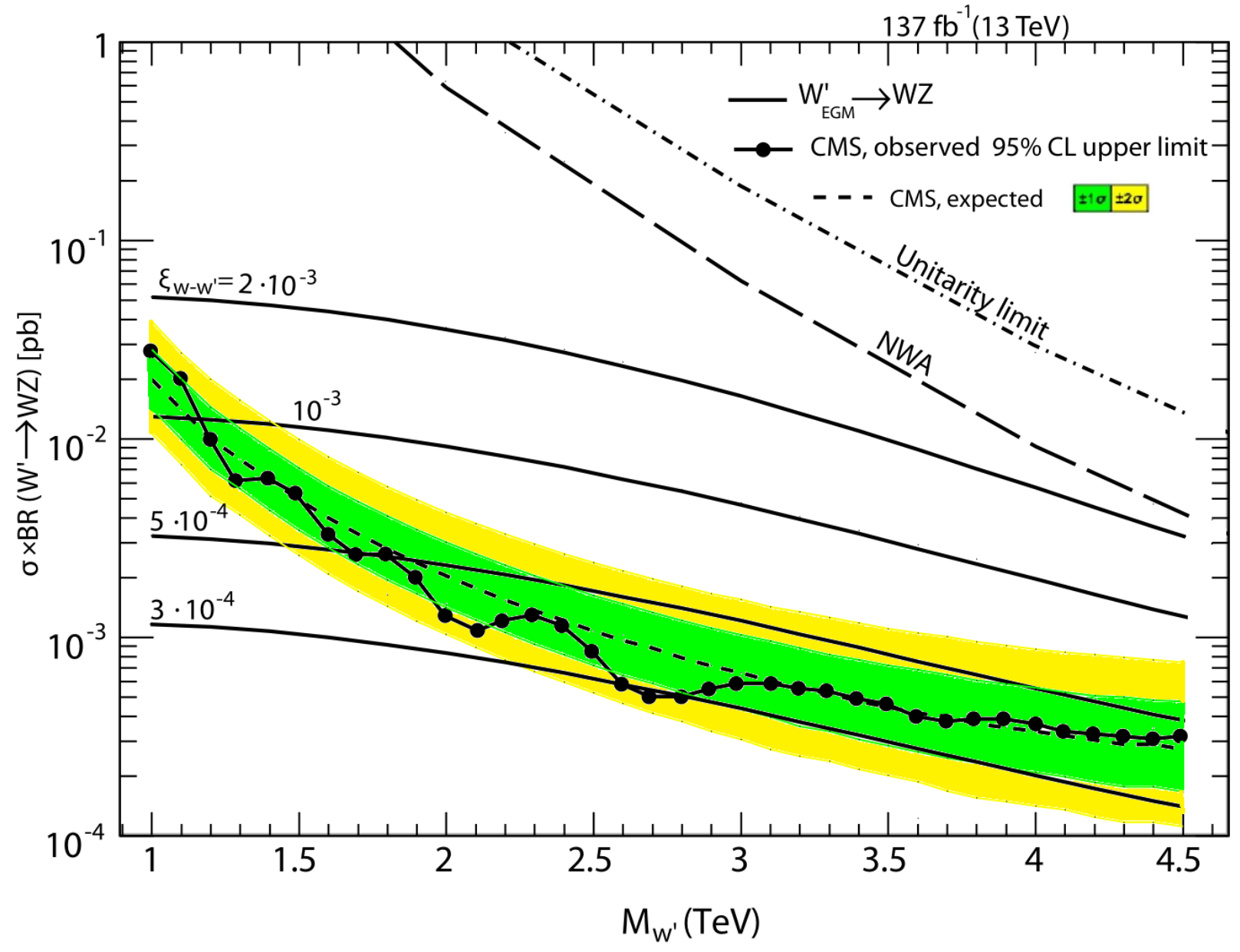}
\includegraphics[scale=0.52]{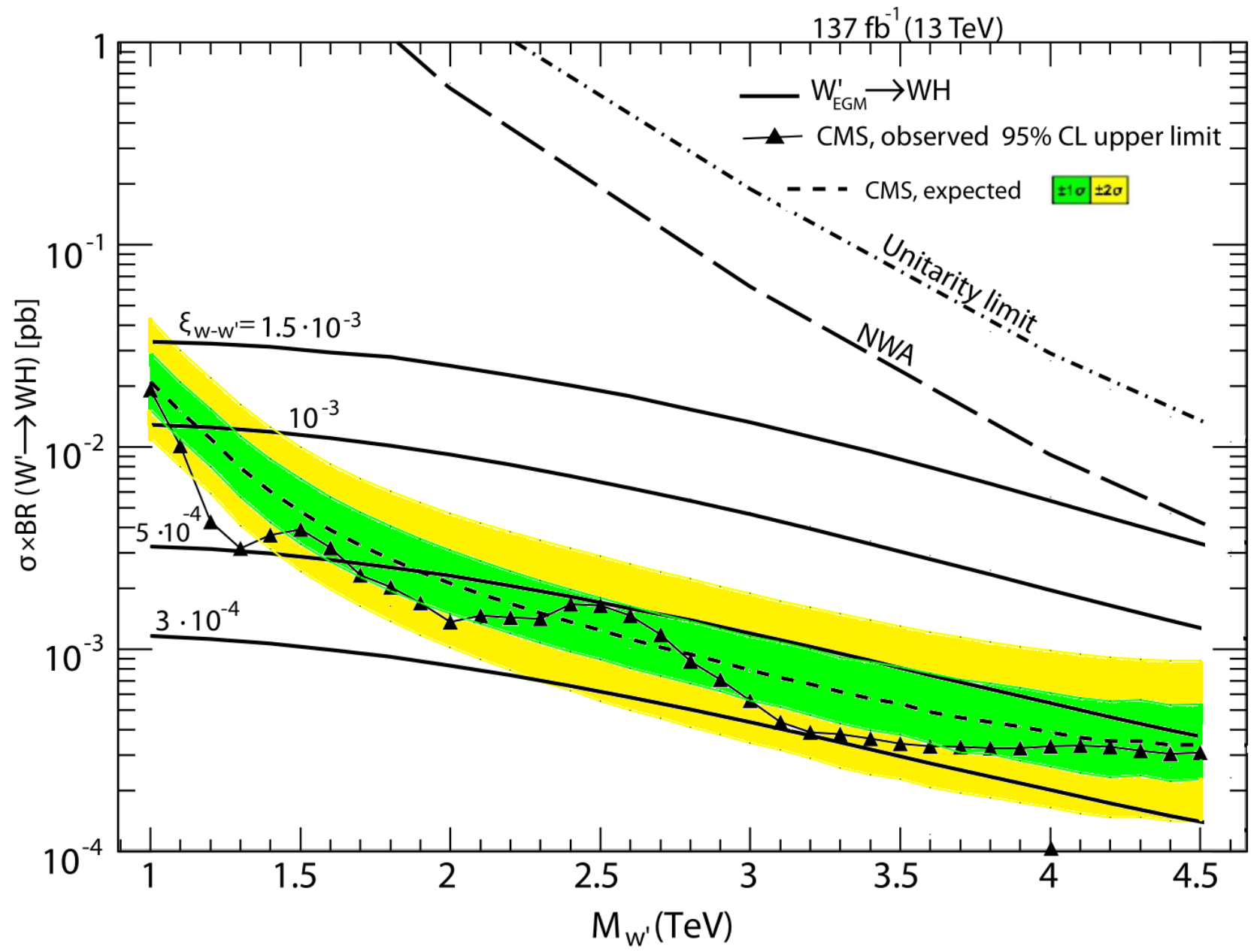}
\end{center}
\caption{
Left panel: Observed and expected $95\%$ C.L. upper limits on $\sigma_{95\%}\times \text{BR}(W'\to WZ)$, showing CMS data on the semileptonic final states for  $137~\text{fb}^{-1}$
\cite{CMS:2021klu,hepdata102645}. The inner (green) and outer (yellow) bands around the expected limits representing $\pm 1\sigma$ and $\pm 2\sigma$ uncertainties, respectively, as determined by CMS.
The theoretical production cross sections $\sigma(pp\to W'X)\times \text{BR}(W'\to
WZ)$ for the ${\rm EGM}$ are shown by solid curves with mixing factors $\xi_{W\text{-}W^\prime}$ attached to the curves \cite{Osland:2020onj}.  The NWA and unitarity constraints are also shown \cite{Alves:2009aa, Serenkova:2019zav}.
Right panel: Same as in the left panel but for the process $pp\to W'\to WH$ \cite{CMS:2021fyk,hepdata101374}.
}
\label{Fig:cross_sect_Wprime}
\end{figure}

In Fig.~\ref{Fig:cross_sect_Wprime}, we show the observed
$95\%$ C.L. upper limits on the production cross section times the branching
fraction, $\sigma_{95\%}\times \text{BR}(W'\to WZ)$ (left panel) and $\sigma_{95\%}\times \text{BR}(W'\to WH)$ (right panel), as functions of the $W'$ mass.

\begin{figure}[htt]
\begin{center}
\includegraphics[scale=0.52]{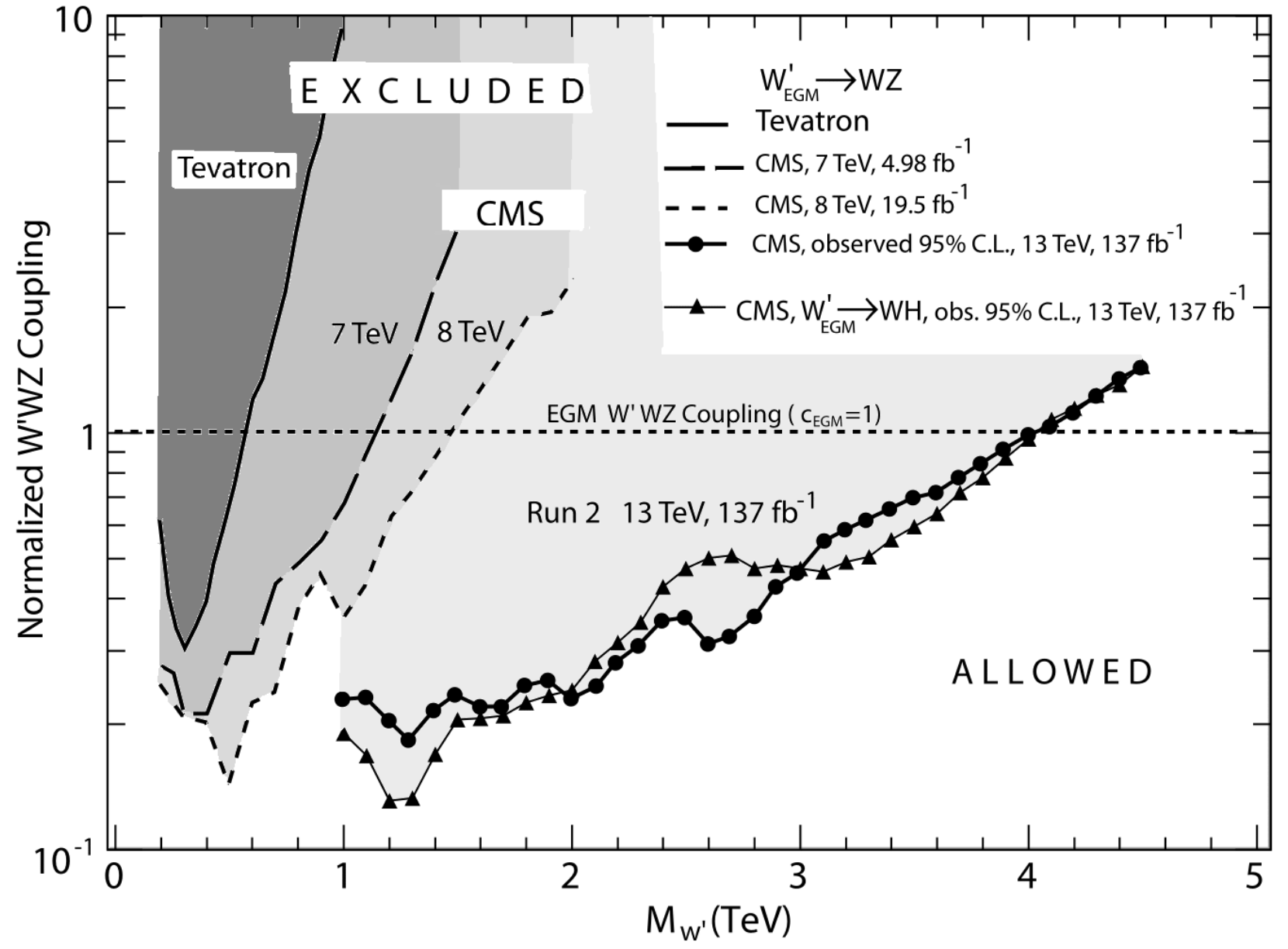}
\includegraphics[scale=0.52]{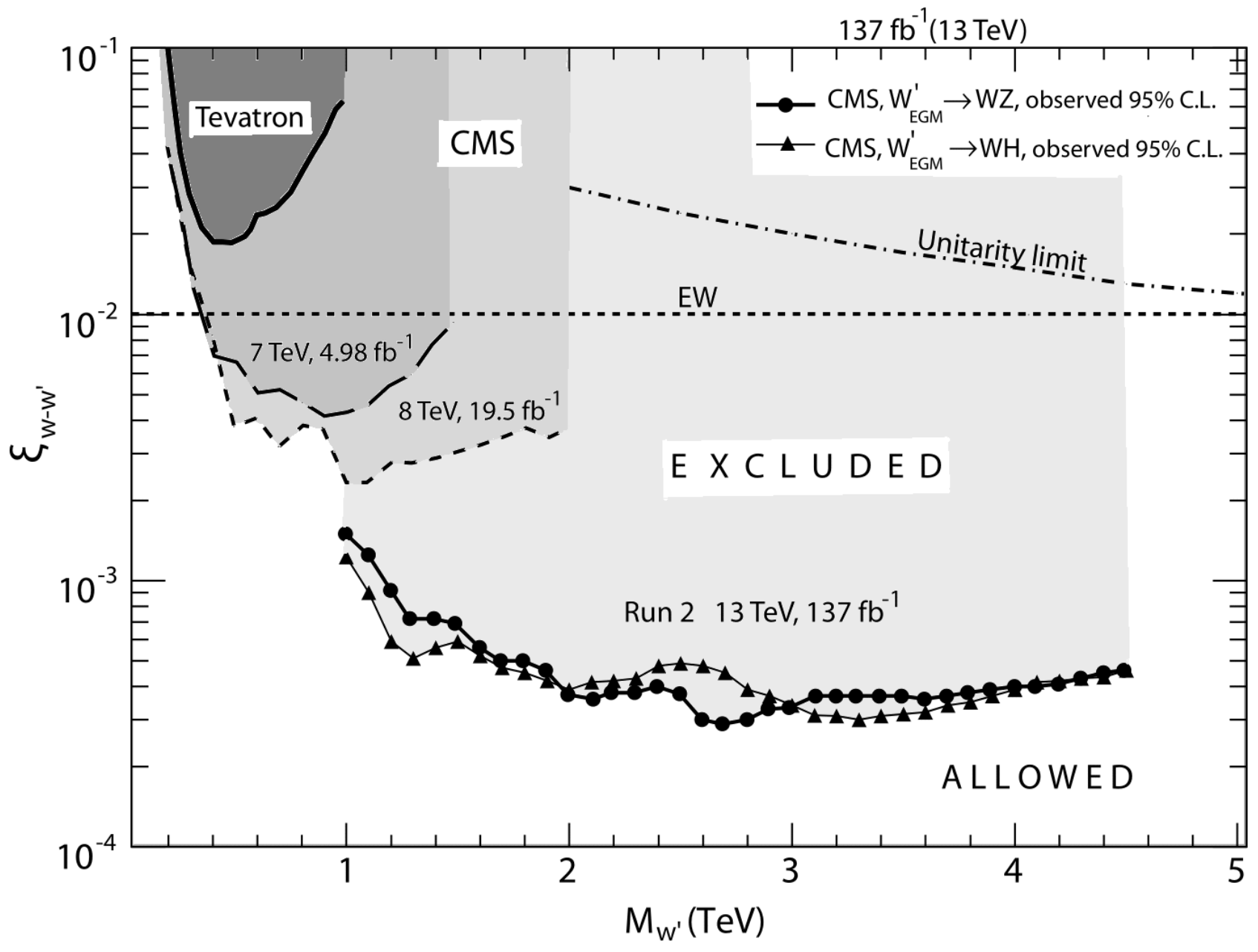}
\end{center}
\caption{Left panel:
Upper limit at 95\% C.L. on the strength of the $g_{W'WZ}$ coupling normalized to the EGM trilinear gauge coupling of $g_{W'WZ}$ for ${\cal C}=1$, as a function of the $W'$ mass,
obtained from  direct search constraints from the Tevatron in
$p\bar{p}\to WZX$ (dark shaded area) as well as from the LHC
searches for $p{p}\to WZX$ at 7 TeV and 8 TeV (Run~1) (respectively, dark and light gray
areas)  and at 13~TeV from diboson production of $W'\to WZ$ and $W'\to WH$ in semileptonic  final state using the  full Run~2 CMS data set. The regions above each curve for the $WZ$ and $WH$  channels are excluded. Right panel: Same as in the left panel but in the ($M_{W'}, \xi_{W-W'}$) plane.
}
\label{Fig:Wprime_constr}
\end{figure}

The theoretical production cross sections multiplied by the branching ratio of $W'$ into $WZ/WH$ bosons, $\sigma (pp\to W'X)\times {\rm BR}(W'\to WZ/WH)$, as functions of the
two parameters ($M_{W'}$, $\xi_{W\text{-}W^\prime}$) can be found in
 \cite{Serenkova:2019zav,Osland:2020onj} and are here compared with the limits established by the CMS experiment, $\sigma_{95\%} \times {\rm BR}(W'\to WZ/WH)$ \cite{CMS:2021klu,hepdata102645}.
The simulation of signals for the EGM $W'$ is based on an adapted version of
the leading-order PYHTHIA 8.2 event generator \cite{Sjostrand:2014zea}.
A mass-dependent $K$ factor is adopted to rescale the LO
PYTHIA prediction to the the NNLO one, using the
ZWPROD \cite{Hamberg:1990np} software.  The factorization and renormalization scales are both set to the $W'$ mass.

The area below the long-dashed curve labelled ``NWA'' corresponds to the region where
the $W'$ resonance width is predicted to be less than 5\% of its
mass, corresponding to the best detector resolution of
the searches, where the narrow-width assumption is satisfied.
We also show a curve labelled  ``Unitarity limit'' that corresponds to the unitarity bound (see, e.g. \cite{Osland:2020onj,Serenkova:2019zav,Alves:2009aa}).

The theoretical curves for the cross sections $\sigma(pp\to W'X)
\times {\rm BR}(W'\to WZ)$, in descending order, correspond to
values of the $W$-$W'$ mixing factor $\xi_{W\text{-}W^\prime}$ from 0.002 to 0.0003. Similarly, $\sigma(pp\to W'X)
\times {\rm BR}(W'\to WH)$ is shown with $\xi_{W\text{-}W^\prime}$ from 0.0015 to 0.0003. The intersection points of the measured upper
limits on the production cross section with these theoretical
cross sections for various values of $\xi_{W\text{-}W^\prime}$ give the corresponding upper
bounds on $\xi_{W\text{-}W^\prime}$, displayed  in
Fig.~\ref{Fig:Wprime_constr}.

From the comparison of sensitivities of the processes (\ref{wz}) and (\ref{wh}) to $W'$ with different decay channels illustrated in Fig.~\ref{Fig:Wprime_constr},  one can  conclude that
the sensitivity of the $W'\to WZ$ channel is quite comparable with that for the
$W'\to WH$ channel within the whole range of the $W'$ mass, from 1.0 TeV to 4.5 TeV. These features are illustrated in Fig.~\ref{Fig:Wprime_constr}.
Notice that the two panels of Fig.~\ref{Fig:Wprime_constr} provide the same information.
The different types of plots presented in those panels based on the specific parametrization of the $W$-$W'$ mixing employed there, namely expressed in terms of a normalized $g_{W'WZ}$ coupling \cite{CMS:2012foj,CMS:2014jem,ATLAS:2013lti}  (left panel) and, alternatively in terms of the mixing parameter $\xi_{W-W'}$ (right panel).  Throughout the paper we use the latter one proposed by CDF \cite{Aaltonen:2010ws} and exploited in previous analyses of ATLAS data sets \cite{Serenkova:2019zav,Pankov:2019yzr,Osland:2020onj} .

For reference, we display limits on the $W'$ parameters
from the Tevatron (CDF) \cite{Aaltonen:2010ws} as well as from CMS
obtained at 7 and 8 TeV of LHC data taking in Run~1  \cite{Serenkova:2019zav,Osland:2020onj}.
Fig.~\ref{Fig:Wprime_constr} (right panel) shows that the CDF experiment at the Tevatron \cite{Aaltonen:2010ws} excludes EGM $W'$ bosons with $\xi_{W\text{-}W^\prime}\gsim 2\cdot 10^{-2}$ in the resonance mass range 0.25~TeV $<M_{W'}<$ 1~TeV at the $95\%$ C.L., whereas CMS at LHC in Run~1 improved those constraints, excluding  $W'$ boson parameters at $\xi_{W\text{-}W^\prime}\gsim 2\cdot 10^{-3}$ in the mass range 0.2~TeV  $<M_{W'}<$ 2~TeV.

As expected, the increase of the time-integrated luminosity up to
137~fb$^{-1}$ leads to increased sensitivity of the diboson channels under study over the whole mass range of 1.0~TeV $<M_{W'}<$ 4.5~TeV and allows to set stronger constraints on the mixing angle $\xi_{W\text{-}W^\prime}$, excluding $\xi_{W\text{-}W^\prime} >  2.5\cdot 10^{-4}$  as shown in Fig.~\ref{Fig:Wprime_constr} (right panel).
Our results extend the sensitivity much beyond the corresponding CDF Tevatron  results \cite{Aaltonen:2010ws}  as well as the ATLAS and CMS sensitivity attained
at 7 and 8~TeV. Also, for the first time, we set $W'$ limits as functions
of the mass $M_{W'}$ and mixing factor $\xi_{W\text{-}W^\prime}$ from the study of the diboson production and subsequent decay into semileptonic final states at the LHC at 13~TeV
with the full CMS Run~2 data set.
The exclusion region obtained in this way on the
parameter space of the $W'$ naturally supersedes the corresponding exclusion area obtained for
time-integrated luminosity of 36.1~fb$^{-1}$ at CMS in the semileptonic channel as reported in \cite{Osland:2017ema}. The limits on the $W'$ parameters presented in this section obtained from the diboson $WZ$ and $WH$ production in semileptonic final states, corresponding to  a time-integrated luminosity of 137~fb$^{-1}$, are quite complementary to those obtained with the entire ATLAS Run~2
 data set \cite{Osland:2020onj} and the best to date.

\section{Hadron  production and decay of $Z'$ boson}
\label{sect:productionZp}

We shall next consider $Z'$ boson production in $pp$ collision and its subsequent decay into diboson channels, $Z^\prime \to W^+W^-$ and $Z^\prime \to ZH$. Specificlly, we concentrate on the models with extended gauge sector predicting the existence of $Z'$ bosons, such as $E_6$, left-right symmetric LR and the EGM. The processes under study are:
\begin{subequations}
\begin{eqnarray}
\sigma(pp\to Z' X\to W^+W^-\,X)&  = & \sigma(pp\to Z'\,X) \times \text{BR}(Z' \to
W^+W^-)\;\label{sigww}, \\
\sigma(pp\to Z' X\to ZH\,X)&  = & \sigma(pp\to Z'\,X) \times \text{BR}(Z' \to
ZH)\;\label{sigzh}.
\end{eqnarray}
\end{subequations}
Here, $\sigma(pp\to Z' X)$ is the  total (theoretical) $Z'$ production cross section,
$\text{BR}(Z' \to W^+W^-)=\Gamma_{Z'}^{WW}/\Gamma_{Z'}$
and
$\text{BR}(Z' \to ZH)=\Gamma_{Z'}^{ZH}/\Gamma_{Z'}$
with
$\Gamma_{Z'}$ the total width of the $Z'$.
Among the $Z'$ models, we start out with a  discussion of the EGM.

\subsection{The $Z'$ width}
\label{sect:Zprime-width}

In the computation of the total width $\Gamma_{Z'}$ we take into account  the
following channels: $Z'\to f\bar f$, $W^+W^-$, and $ZH$
\cite{Salvioni:2009mt,Gulov:2018zij,Pankov:2019yzr,Osland:2020onj}, where $H$ is the SM Higgs boson and $f$ refers to the SM fermions ($f=l,\nu,q$). Throughout the paper we shall
ignore the couplings of the $Z'$ to any beyond-SM particles such as
right-handed neutrinos, as well as to
SUSY partners and any other exotic fermions.
Any additional states may increase the width of the $Z'$.

The total width $\Gamma_{Z'}$ of the $Z'$ boson can then be written as
follows:
\begin{equation}\label{gamma2}
\Gamma_{Z'} = \sum_f \Gamma_{Z'}^{ff} + \Gamma_{Z'}^{WW} +
\Gamma_{Z'}^{ZH}.
\end{equation}
The two last terms are due to $Z$-$Z'$
mixing.  For the range of $M_{Z'}$
values below $\sim 3$ TeV,  the dependence of
$\Gamma_{Z'}$ on the values of $\xi_{Z\text{-}Z'}$ (within its allowed range) is unimportant.
Therefore, in this mass range, one can approximate the total width as $\Gamma_{Z'} \approx \sum_f
 \Gamma_{Z'}^{ff}$, where the sum runs over SM fermions only.
 Within the approximation above, one can quantify the ratio of $\Gamma_{Z'}/M_{Z'}$ for the benchmark EGM as $3\%$, whereas for the $\psi,\,\eta,\,\chi$ and
${\rm LR}$ models it varies from $0.5\%$ to $2.0\%$.

 However, for larger $Z'$ masses, $M_{Z'}>4$ TeV, there is an
enhancement in the coupling that cancels the suppression due to the tiny $Z$-$Z'$
mixing parameter $\xi_{Z\text{-}Z'}$ \cite{Salvioni:2009mt}. We note that the
``Equivalence theorem'' \cite{Chanowitz:1985hj} suggests a
value for $\text{BR}(Z'\to ZH)$ comparable to $\text{BR}(Z'\to W^+W^-)$,
 up to electroweak symmetry breaking effects and phase-space factors.
 Throughout this paper, for definiteness, we adopt a scenario where
both partial widths are comparable, $\Gamma_{Z'}^{ZH}\simeq
\Gamma_{Z'}^{WW}$ for heavy $M_{Z'}$
\cite{Barger:1987xw,Barger:2009xg,Dib:1987ur}.

For all $M_{Z'}$ values of interest for our analysis the width
of the $Z'$ boson is considerably smaller than the experimental
mass resolution $\Delta M$. We adopt the approximation $\Delta M/M\approx 5\% $,
as reported, e.g., in
\cite{Sirunyan:2017nrt} for reconstructing the diboson invariant mass of the $W^+W^-$ and $ZH$
systems.

\begin{figure}[htb]
\begin{center}
\includegraphics[scale=0.5]{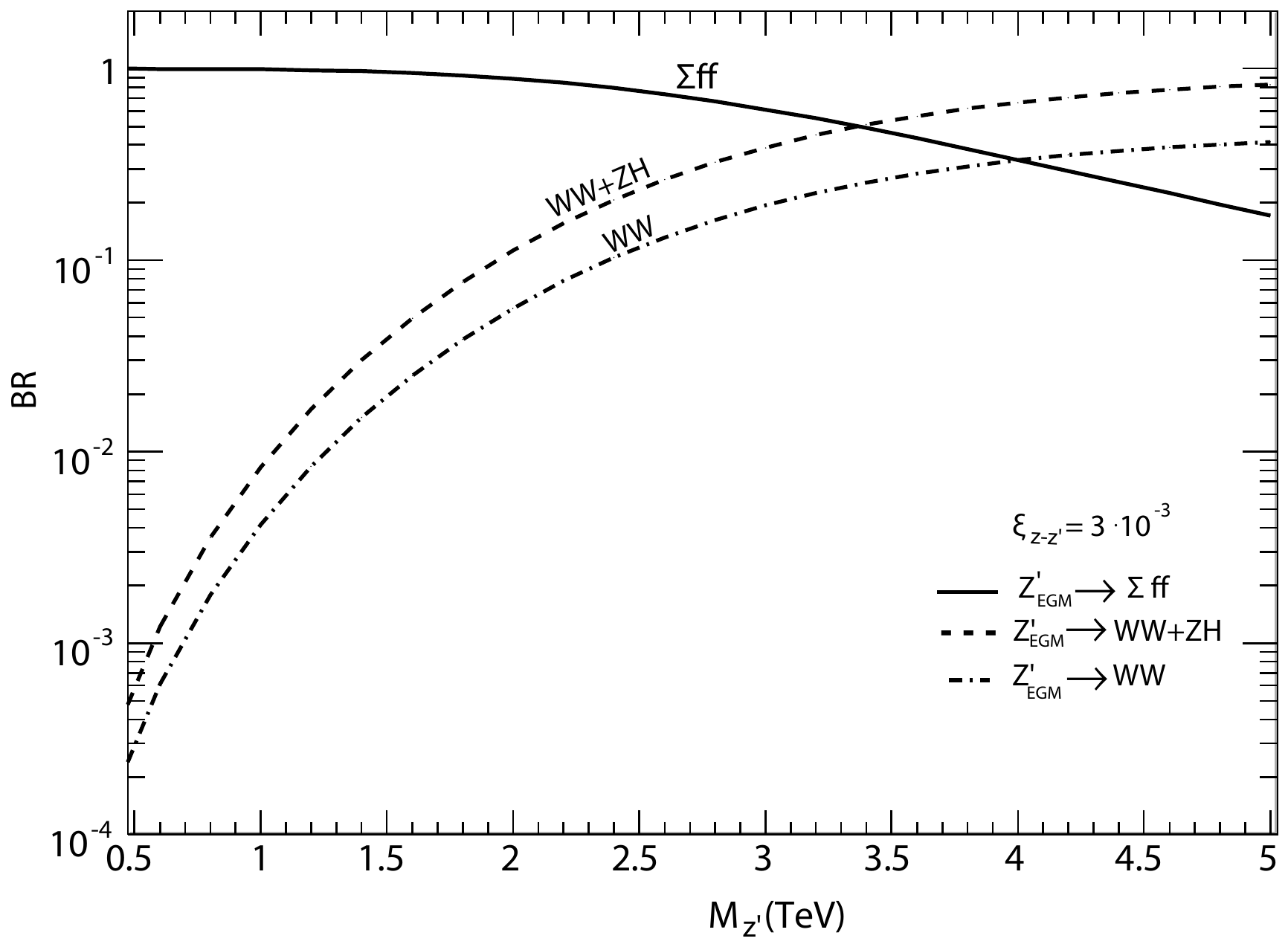}
\end{center}
\caption{
Branching ratios $\text{BR}(Z'\to\sum{f\bar{f}})$ (solid), $\text{BR}(Z'\to W^+W^-)$ (dash-dotted), and $\text{BR}(Z'\to W^+W^-+ZH)$ (dashed) vs $M_{Z'}$ in the EGM for a $Z$-$Z'$ mixing factor
$\xi_{Z\text{-}Z^\prime}=3\cdot 10^{-3}$.
It is assumed that $\text{BR}(Z'\to W^+W^-)=\text{BR}(Z'\to ZH)$.
}\label{br-Zprime_egm}
\end{figure}

The partial width of the $Z'\to W^+W^-$ decay
channel can be written as \cite{Altarelli:1989ff}:
\begin{equation}
\Gamma_{Z'}^{WW}=\frac{\alpha_{\rm em}}{48}\cot^2\theta_W\, M_{Z'}
\left(\frac{M_{Z'}}{M_W}\right)^4\left(1-4\,\frac{M_W^2}{M_{Z'}^2}\right)^{3/2}
\left[ 1+20 \left(\frac{M_W}{M_{Z'}}\right)^2 + 12
\left(\frac{M_W}{M_{Z'}}\right)^4\right]\cdot\xi_{Z\text{-}Z'}^2. \label{GammaWW}
\end{equation}
For a fixed mixing factor $\xi_{Z\text{-}Z'}$ and at large $M_{Z'}$ where
$\Gamma_{Z'}^{WW}$ dominates over $\sum_f \Gamma_{Z'}^{ff}$ the total width increases
rapidly with the mass $M_{Z'}$ because of the quintic
dependence of the $W^+W^-$ mode on the $Z'$ mass. In this case, the $W^+W^-$ mode (together with $Z'\to ZH$) becomes dominant and $\text{BR}(Z' \to W^+W^-)\to 0.5$ (this value arises from the assumption $\Gamma_{Z'}^{ZH}=\Gamma_{Z'}^{WW}$), while the fermionic decay channels ($\Gamma_{Z'}^{ff}\propto M_{Z'}$) are increasingly suppressed. These features are illustrated in Fig.~\ref{br-Zprime_egm}, where we plot $\text{BR}(Z'\to W^+W^-)$, $\text{BR}(Z'\to W^+W^-+ZH)$ and $\text{BR}(Z'\to\sum_f\bar ff)$ vs $M_{Z'}$ for the $Z'_{\rm EGM}$ taking $\xi_{Z\text{-}Z^\prime}=3\cdot 10^{-3}$ as a representative case.

\subsection{Constraints on $Z$-$Z'$ mixing and  $M_{Z^\prime}$}
\label{sect:Zprime-lim}

\begin{figure}[htb]
\begin{center}
\includegraphics[scale=0.52]{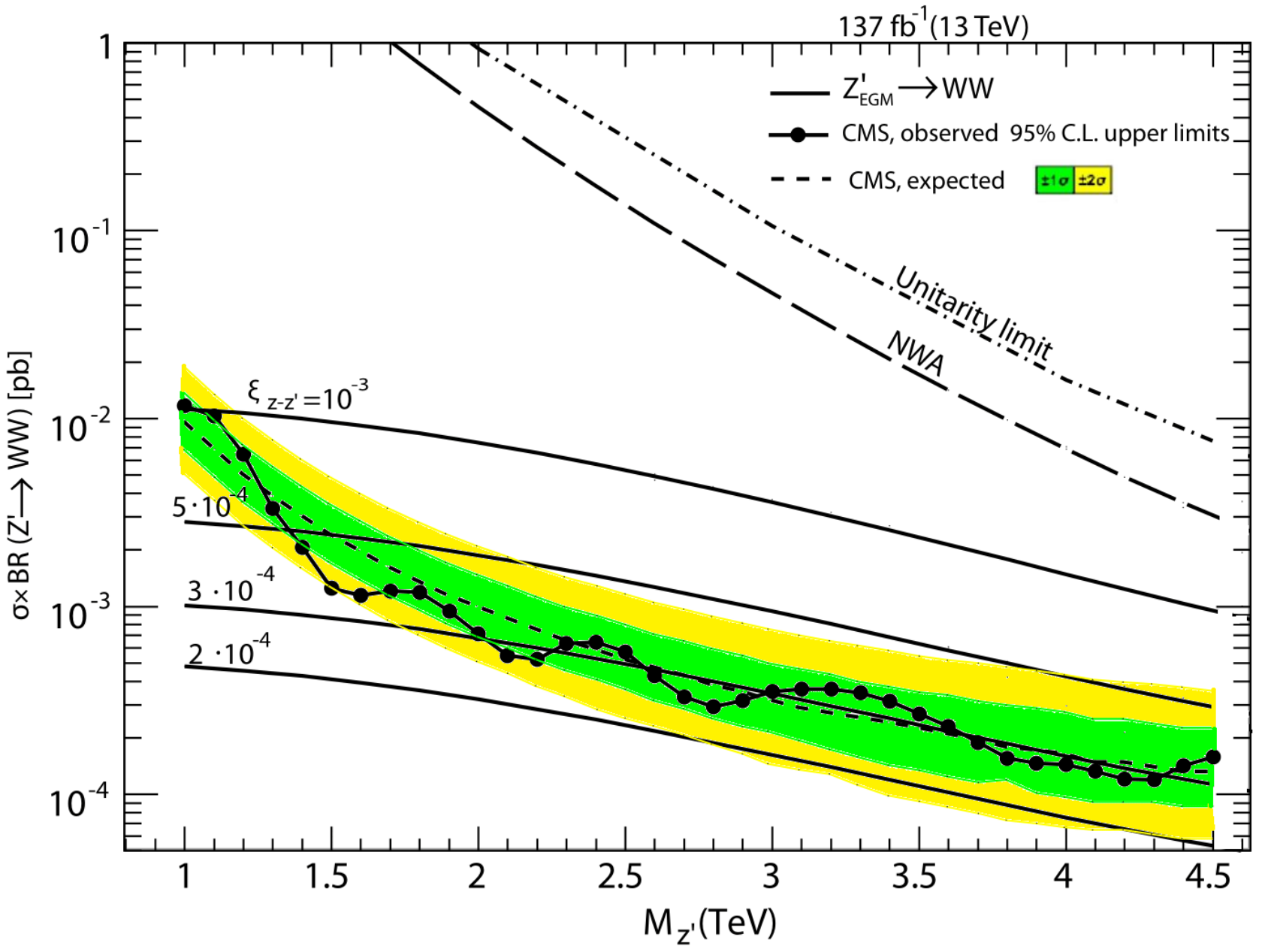}
\includegraphics[scale=0.52]{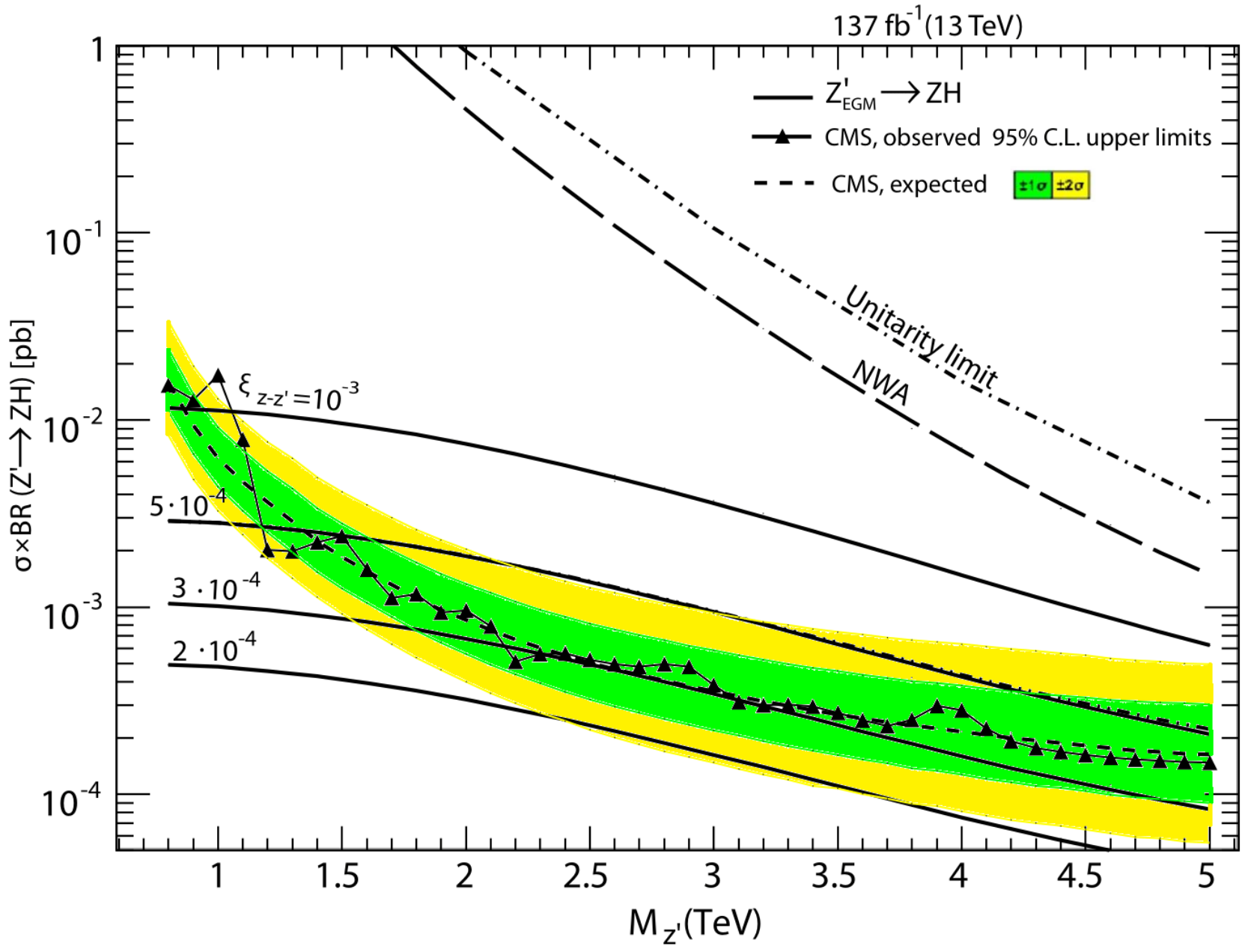}
\end{center}
\caption{
Left panel:
$95\%$ C.L. upper limits on $\sigma_{95\%}\times \text{BR}(Z'\to WW)$,
showing CMS data on the semileptonic final states for  $137~\text{fb}^{-1}$
\cite{CMS:2021klu,hepdata102645}.
The theoretical production cross sections $\sigma(pp\to Z'X)\times \text{BR}(W'\to
WW)$ for the ${\rm EGM}$ are calculated from PYTHIA with a $Z'$
mass-dependent $K$-factor, given by solid curves, for mixing factor $\xi_{Z\text{-}Z^\prime}$
ranging from $10^{-3}$ and down to $2\cdot 10^{-4}$.  The NWA and
unitarity constraints are also shown.
Right panel: Same as in the left panel but for the process $pp\to Z'\to ZH$ \cite{CMS:2021fyk,hepdata101374} .
}
\label{Fig:cross_sect_EGM}
\end{figure}

In Fig.~\ref{Fig:cross_sect_EGM}, we consider the full CMS Run2 data set of time integrated luminosity of 137 fb$^{-1}$ and show the observed  $95\%$ C.L. upper limits on the production cross section times the branching fraction, $\sigma_{95\%}\times \text{BR}(Z'\to
W^+W^-)$ (left panel) and  $\sigma_{95\%}\times \text{BR}(Z'\to ZH)$ (right panel), as functions of the $Z'$ mass, obtained from the semileptonic \cite{CMS:2021klu,hepdata102645,CMS:2021fyk,hepdata101374}  final state.  These figures allow to  make a comparison of the sensitivities of the data to the $Z$-$Z'$ mixing parameter and new gauge boson mass and they  demonstrate the comparable sensitivity to $Z'$ of the $W^+W^-$ and $ZH$ channels over almost the whole allowed $Z'$ mass range. However,  as can be see from Fig.~\ref{Fig:cross_sect_EGM}, the $Z'$ mass range in the $ZH$ channel is somewhat broader than that of the $W^+W^-$ channel, namely 0.8--5.0 TeV vs 1.0--4.5 TeV, respectively.

Then, for $Z'_{\rm EGM}$ we compute the theoretical LHC production cross section multiplied
by the branching ratios into two $W^\pm$ bosons and into $ZH$, $\sigma(pp\to Z'_{\rm EGM} X) \times {\rm
BR}(Z'_{\rm EGM} \to W^+ W^-)$ \cite{Pankov:2019yzr,Osland:2020onj}  and  $\sigma(pp\to Z'_{\rm EGM} X) \times {\rm BR}(Z'_{\rm EGM}\to ZH)$, as functions of the two parameters
($M_{Z'}$, $\xi_{Z\text{-}Z^\prime}$), and compare them with the limits established by the
CMS experiments, $\sigma_{95\%} \times {\rm BR}(Z'\to W^+ W^-)$ (left panel) \cite{CMS:2021klu,hepdata102645} and $\sigma_{95\%} \times {\rm BR}(Z'\to ZH)$ (right panel)  \cite{CMS:2021fyk,hepdata101374}.

\begin{figure}[htb]
\begin{center}
\includegraphics[scale=0.52]{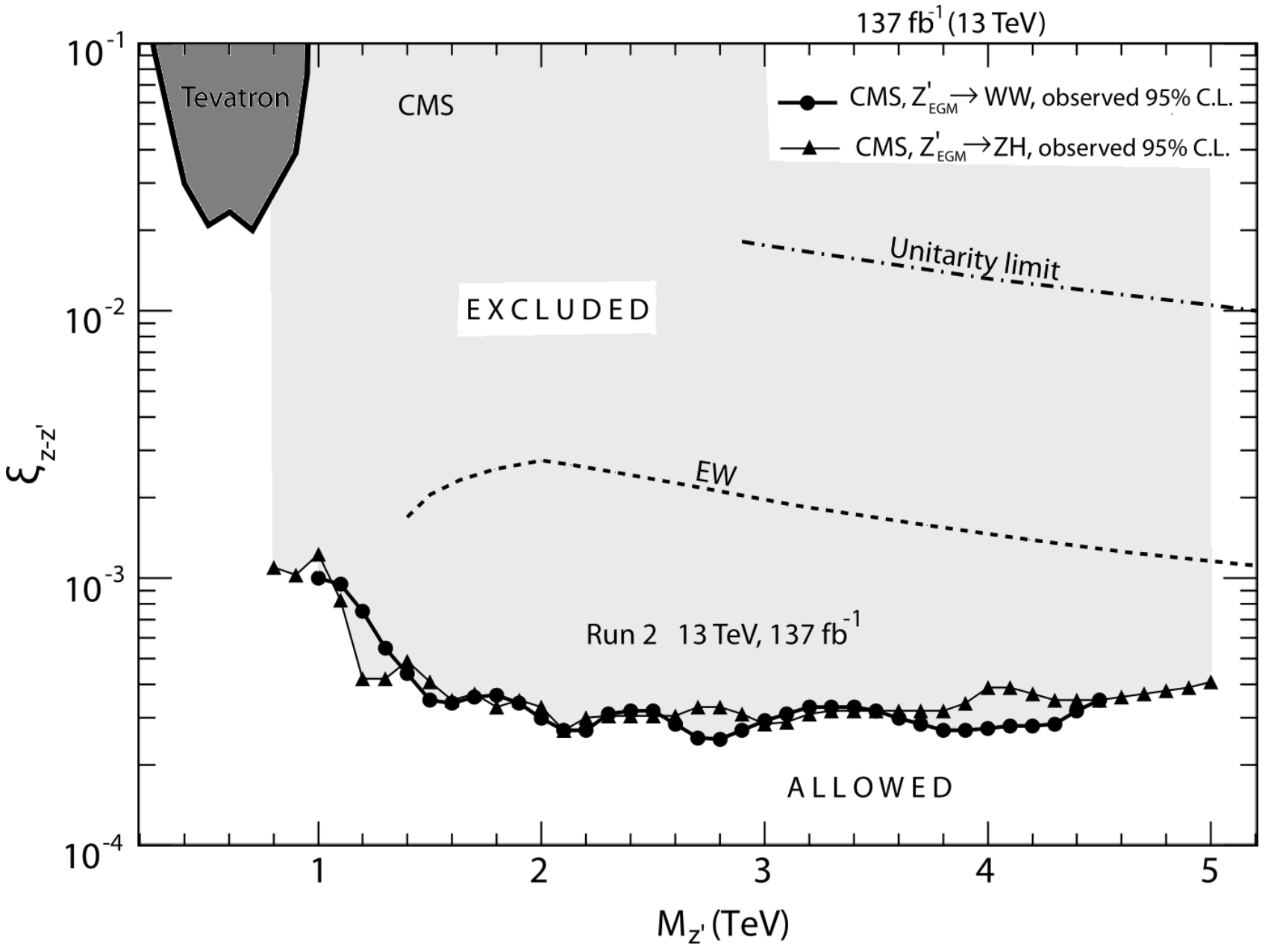}
\end{center}
\caption{
The $Z'_{\rm EGM}$ model: 95\%C.L. exclusion regions in the
two-dimensional ($M_{Z'}$, $\xi_{Z\text{-}Z'}$) plane obtained after
incorporating indirect constraints from electroweak precision data 
(dashed curve labeled ``EW'' \cite{Erler:2009jh}), and
direct search constraints from  the Tevatron in $p\bar{p}\to WZX$ (dark shaded area) \cite{Aaltonen:2010ws} as well as from
the LHC  searches for $pp\to Z'\to WW$ and $pp\to Z'\to ZH$  in semileptonic final states using the full Run~2 CMS data set.
The region above the curves for the $WW$ and $ZH$ channels are excluded.
 }
\label{Fig:bounds-egm}
\end{figure}

The theoretical production cross sections
$\sigma(pp\to Z'_{\rm EGM})\times \text{BR}(Z'_{\rm EGM}\to W^+W^-)$ for $Z'_{\rm EGM}$
boson are calculated from a dedicated modification of PYHTHIA 8.2
\cite{Sjostrand:2014zea}. Higher-order QCD corrections to the signal were
estimated using a $K$-factor, for which we adopt a mass-independent value of
1.9 \cite{Frixione:1993yp,Agarwal:2010sn,Gehrmann:2014fva}.
These theoretical curves for the cross sections, in descending order,
correspond to values of the $Z$-$Z'$ mixing factor $\xi_{Z\text{-}Z^\prime}$
ranging from $10^{-3}$ and down to $2\cdot 10^{-4}$. The NWA and
unitarity constraints are also shown \cite{Pankov:2019yzr,Osland:2020onj,Alves:2009aa}.
The intersection points of the measured
upper limits on the production cross section with this
theoretical cross section for various values of $\xi_{Z\text{-}Z^\prime}$ give the
corresponding bounds on ($M_{Z'}$, $\xi_{Z\text{-}Z^\prime}$), presented
in Fig.~\ref{Fig:bounds-egm}.

Different bounds on the $Z'$ parameter space are collected in Fig.~\ref{Fig:bounds-egm} for the $Z'_{\rm EGM}$ model, showing that at high masses, the limits on $\xi_{Z\text{-}Z^\prime}$ obtained from the full Run~2 data set collected  at $\sqrt{s}=13$ TeV and recorded by the CMS detector
are substantially stronger than that derived from the global analysis
of the precision electroweak data (EW) \cite{Erler:2009jh}, as well as the limits obtained from diboson data at the Tevatron \cite{Aaltonen:2010ws}. Limits obtained separately with CMS from the two channels, $Z'\to W^+W^-$ and $Z'\to ZH$,  are shown for comparison.
It turns out that the diboson channel $Z'\to ZH$, in contrast to  $Z'\to W^+W^-$, allows to place
limits on $Z$-$Z'$ mixing in the  narrow mass ranges such as 0.8~TeV $\leq M_{Z'} \leq$ 1.0~TeV and  4.5~TeV $\leq M_{Z'} \leq$ 5.0~TeV, whereas
in the rest of the resonance mass range, 1.0~TeV $\leq M_{Z'} \leq$ 4.5~TeV, both channels demonstrate  comparable sensitivity to $Z$-$Z'$ mixing.

\begin{figure}[htb]
\begin{center}
\includegraphics[scale=0.52]{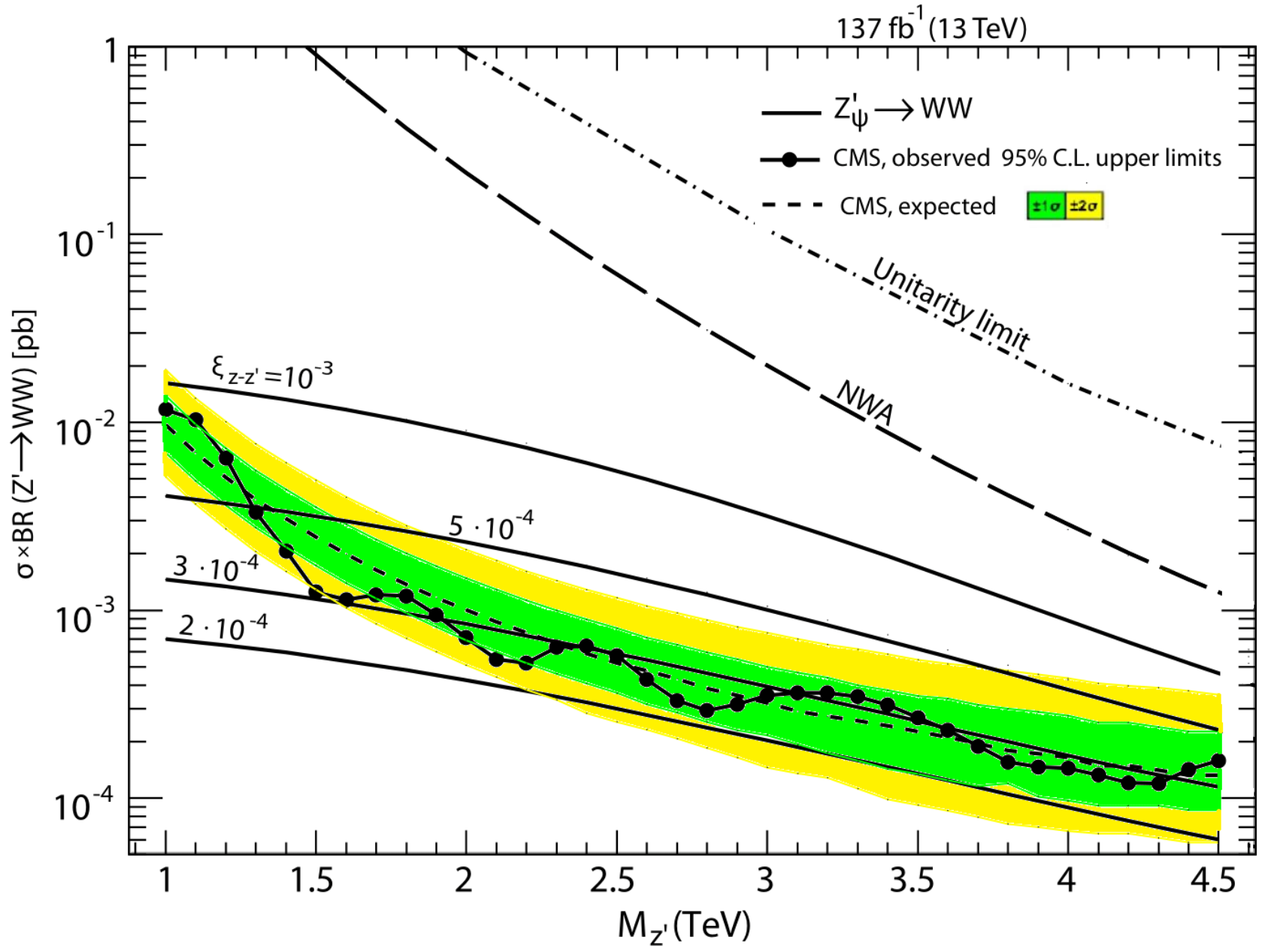}
\includegraphics[scale=0.52]{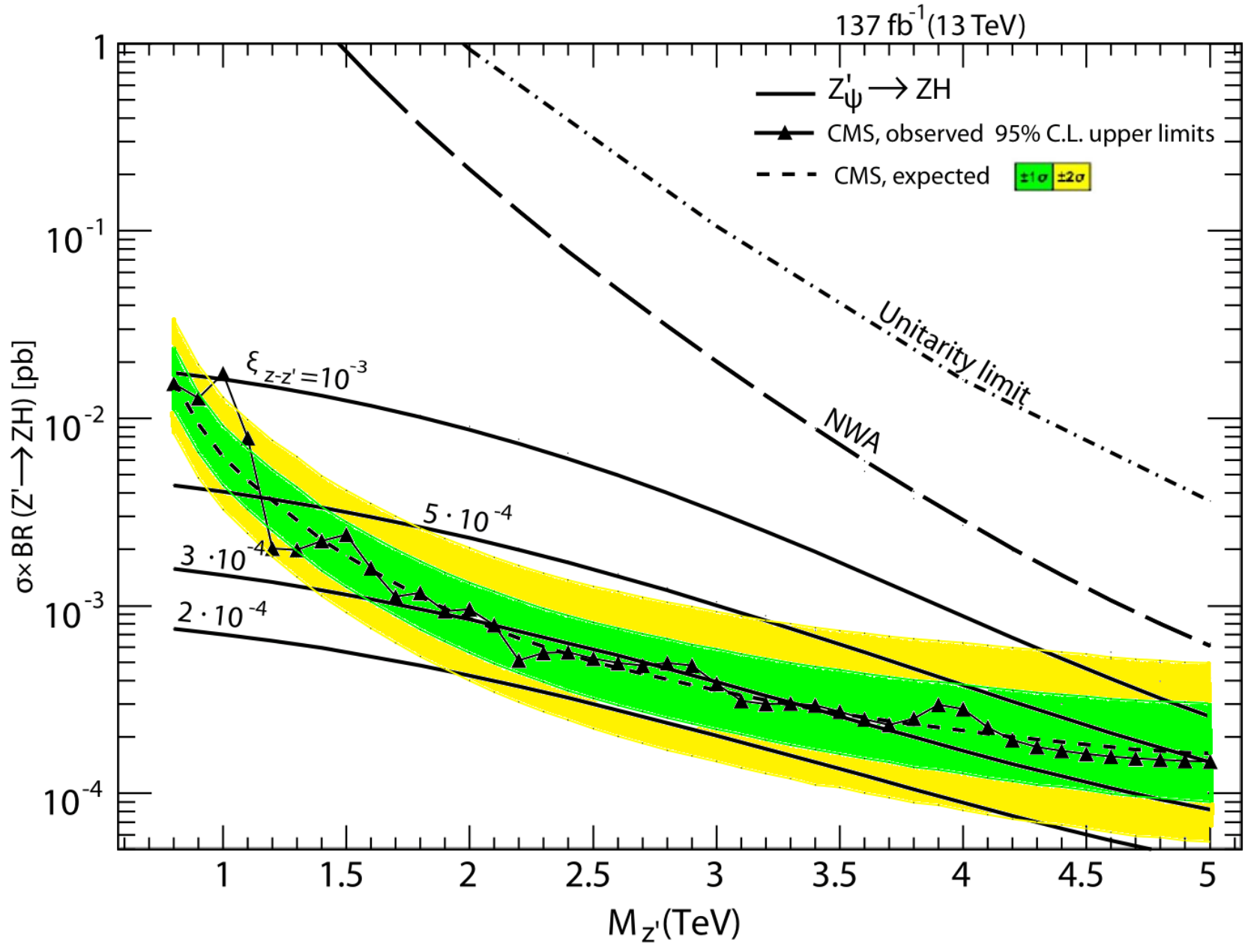}
\includegraphics[scale=0.52]{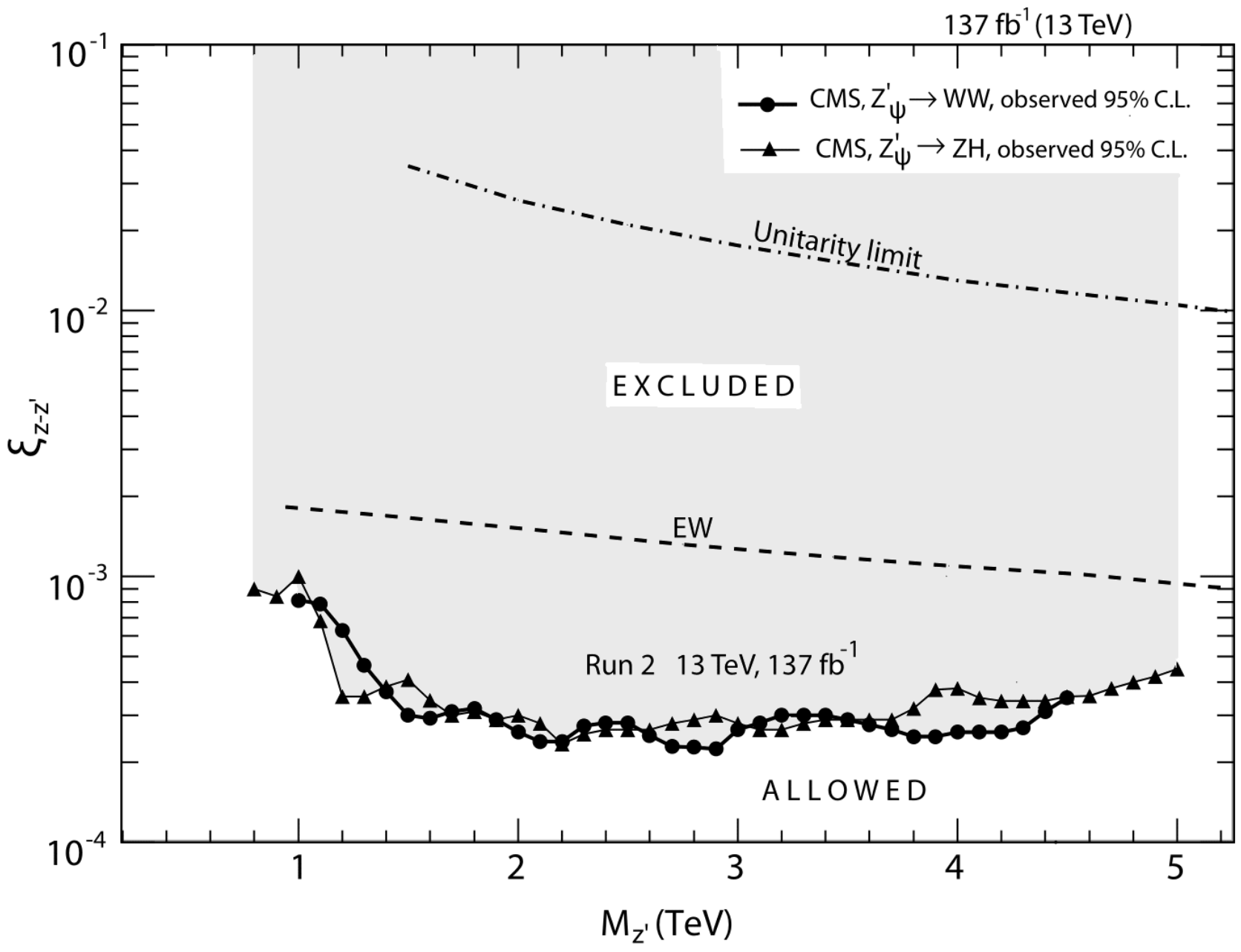})
\end{center}
\caption{
$E_6$ $Z'_{\psi}$ model: Analogous to Figs.~\ref{Fig:cross_sect_EGM}
and \ref{Fig:bounds-egm}, respectively.
}
\label{Fig:psi}
\end{figure}

\begin{figure}[htb]
\begin{center}
\includegraphics[scale=0.5]{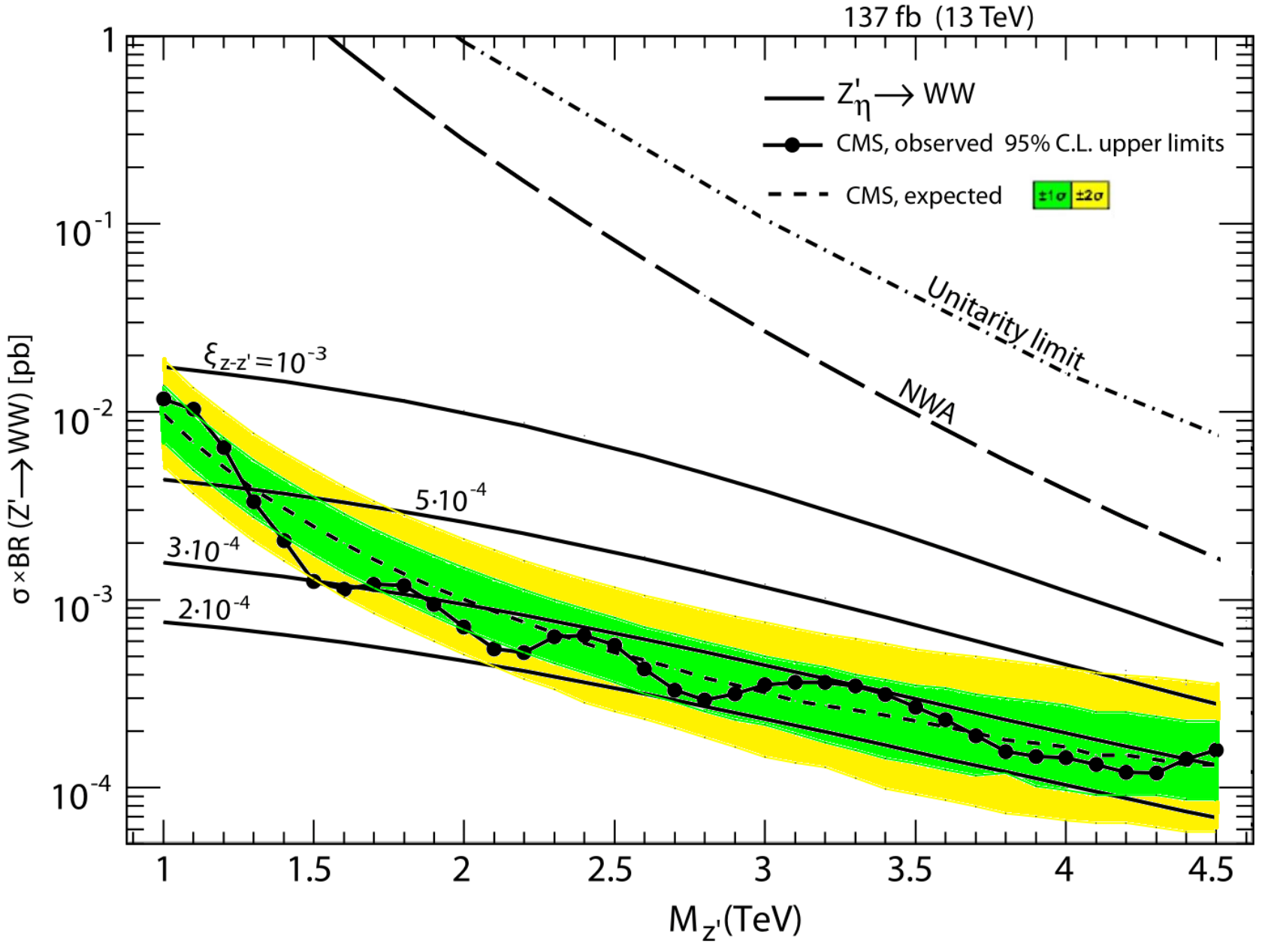}
\includegraphics[scale=0.5]{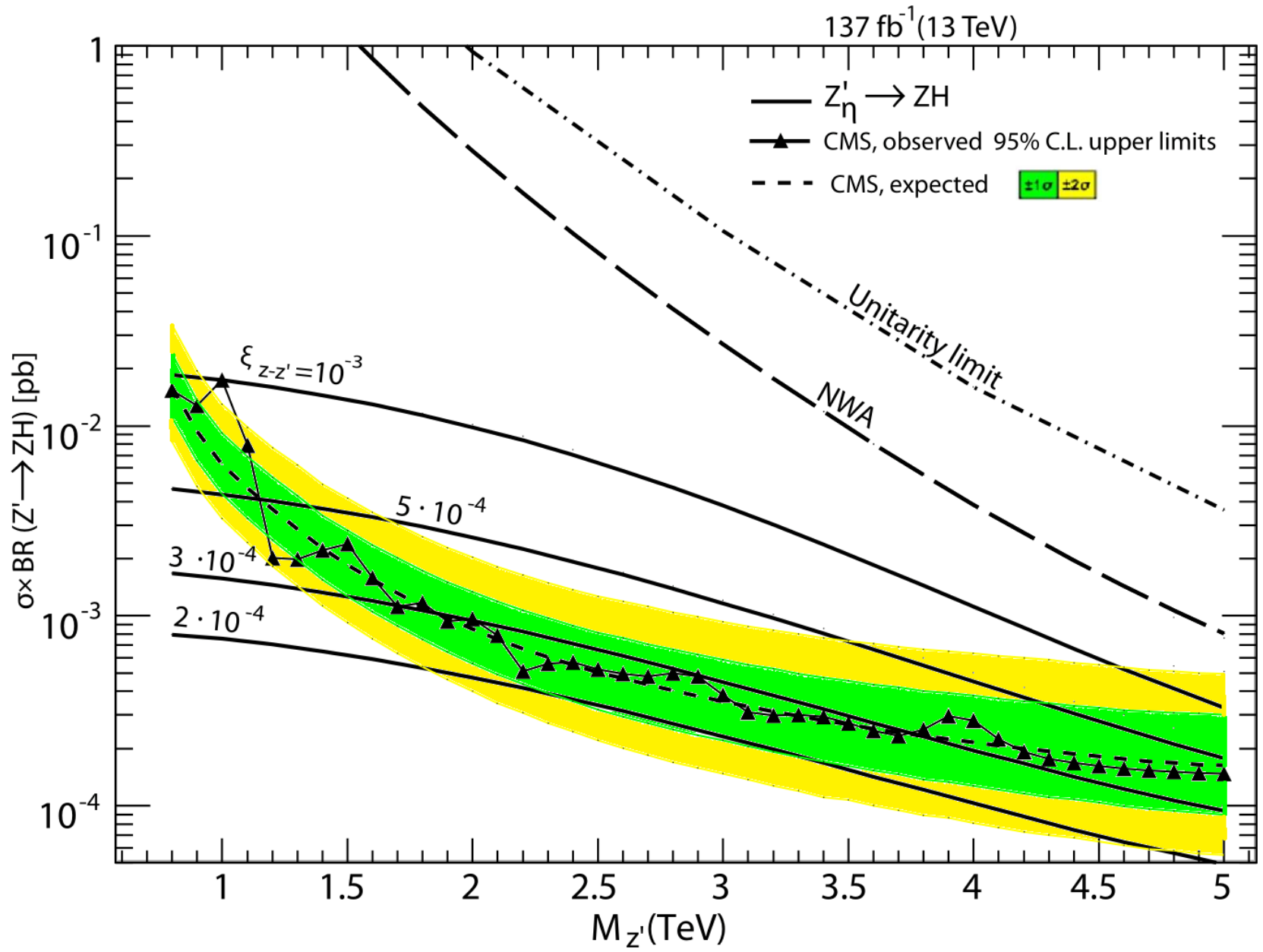}
\includegraphics[scale=0.52]{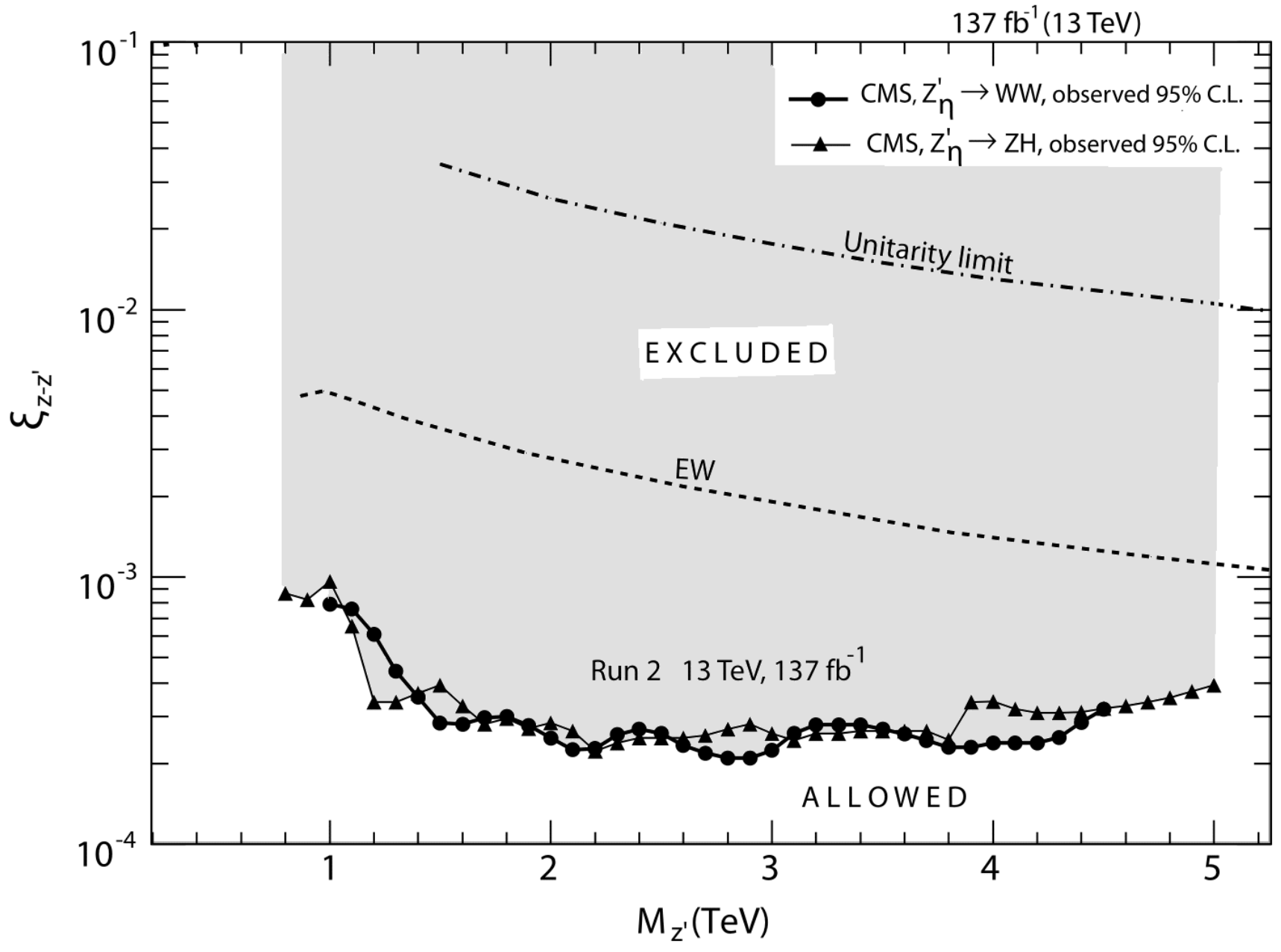}
\end{center}
\caption{
$E_6$ $Z'_{\eta}$ model: Analogous to  Figs.~\ref{Fig:cross_sect_EGM}
and \ref{Fig:bounds-egm}, respectively.
}
\label{Fig:eta}
\end{figure}

\begin{figure}[htb]
\begin{center}
\includegraphics[scale=0.42]{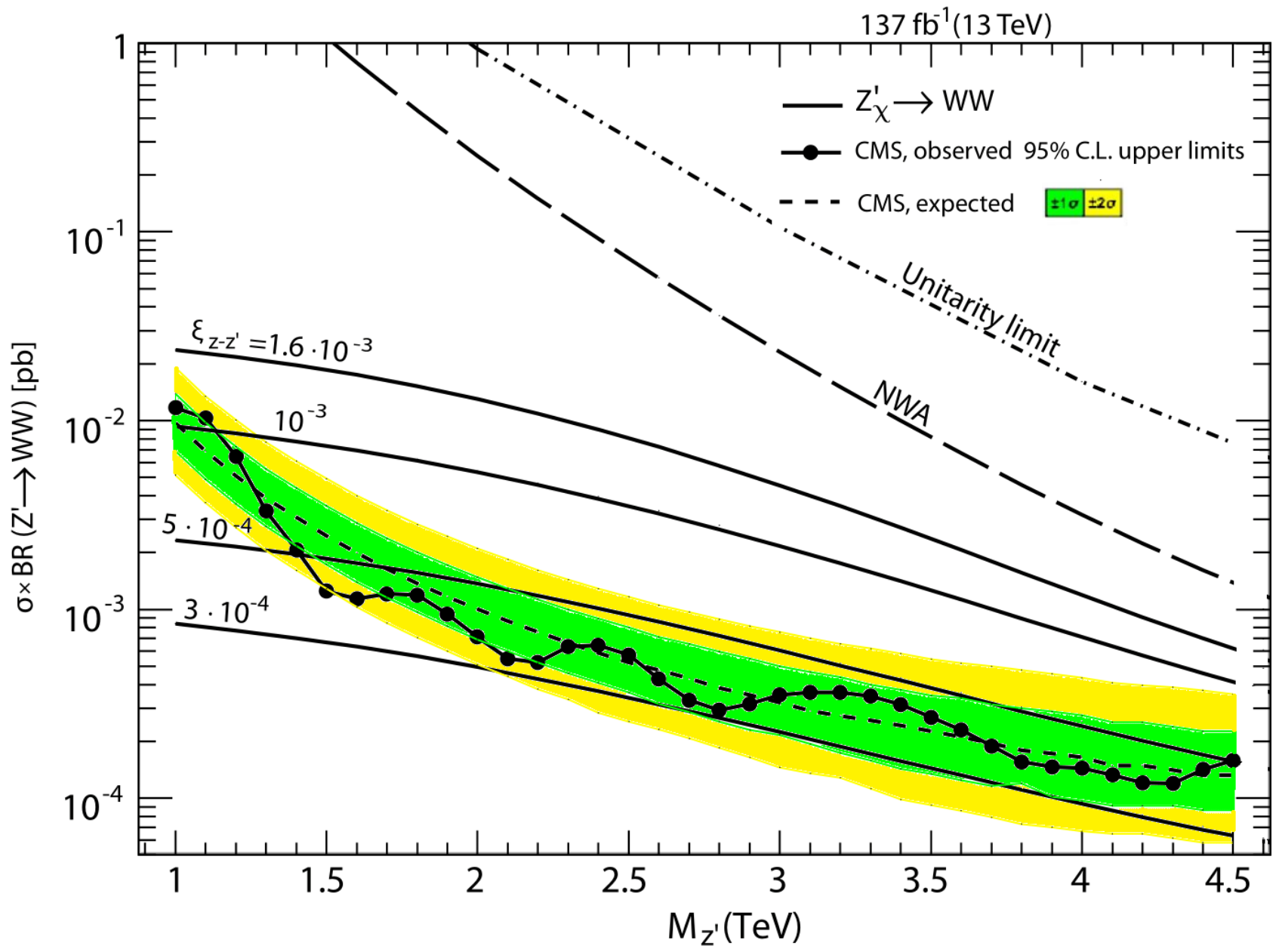}
\includegraphics[scale=0.42]{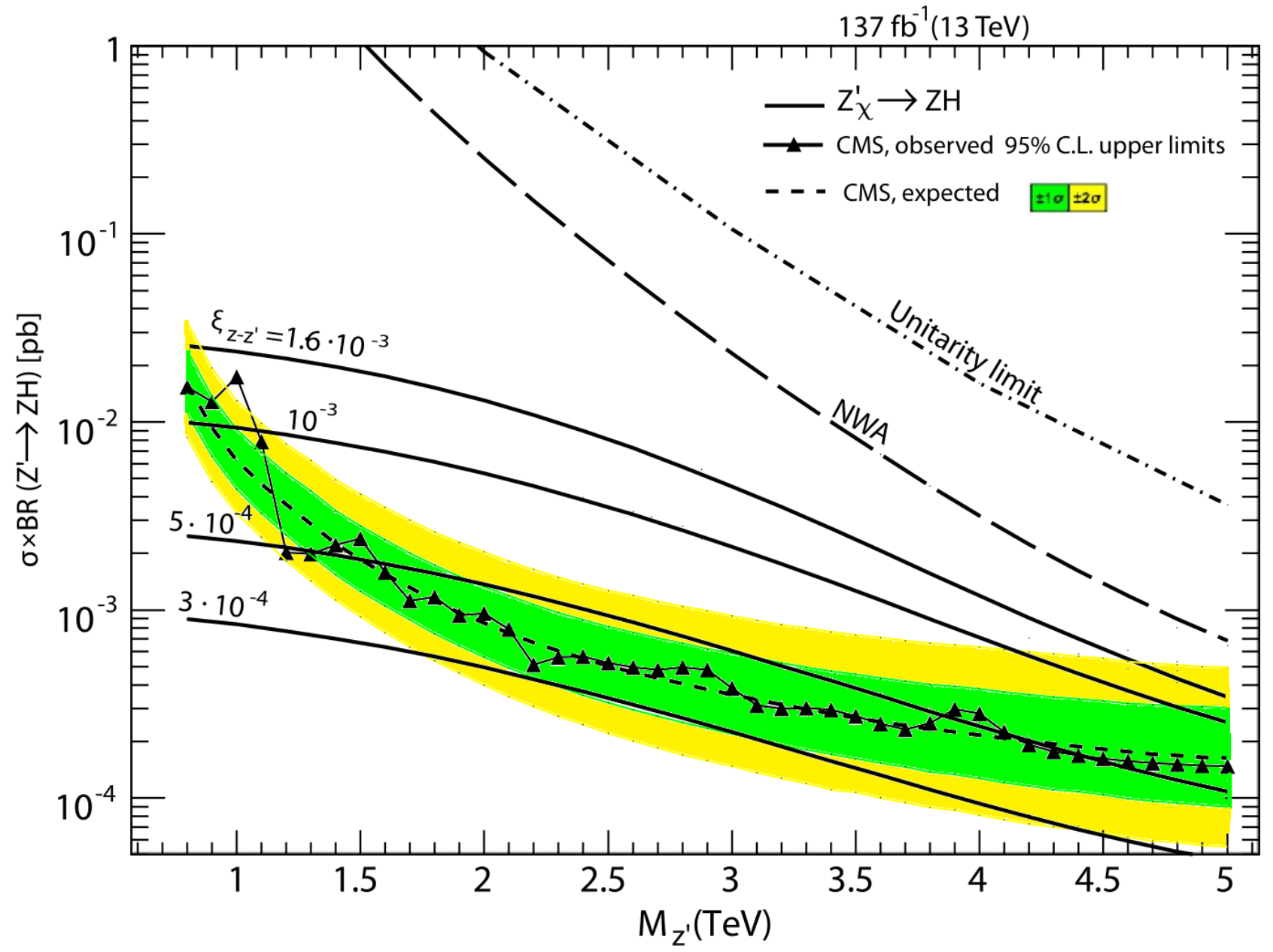}
\includegraphics[scale=0.52]{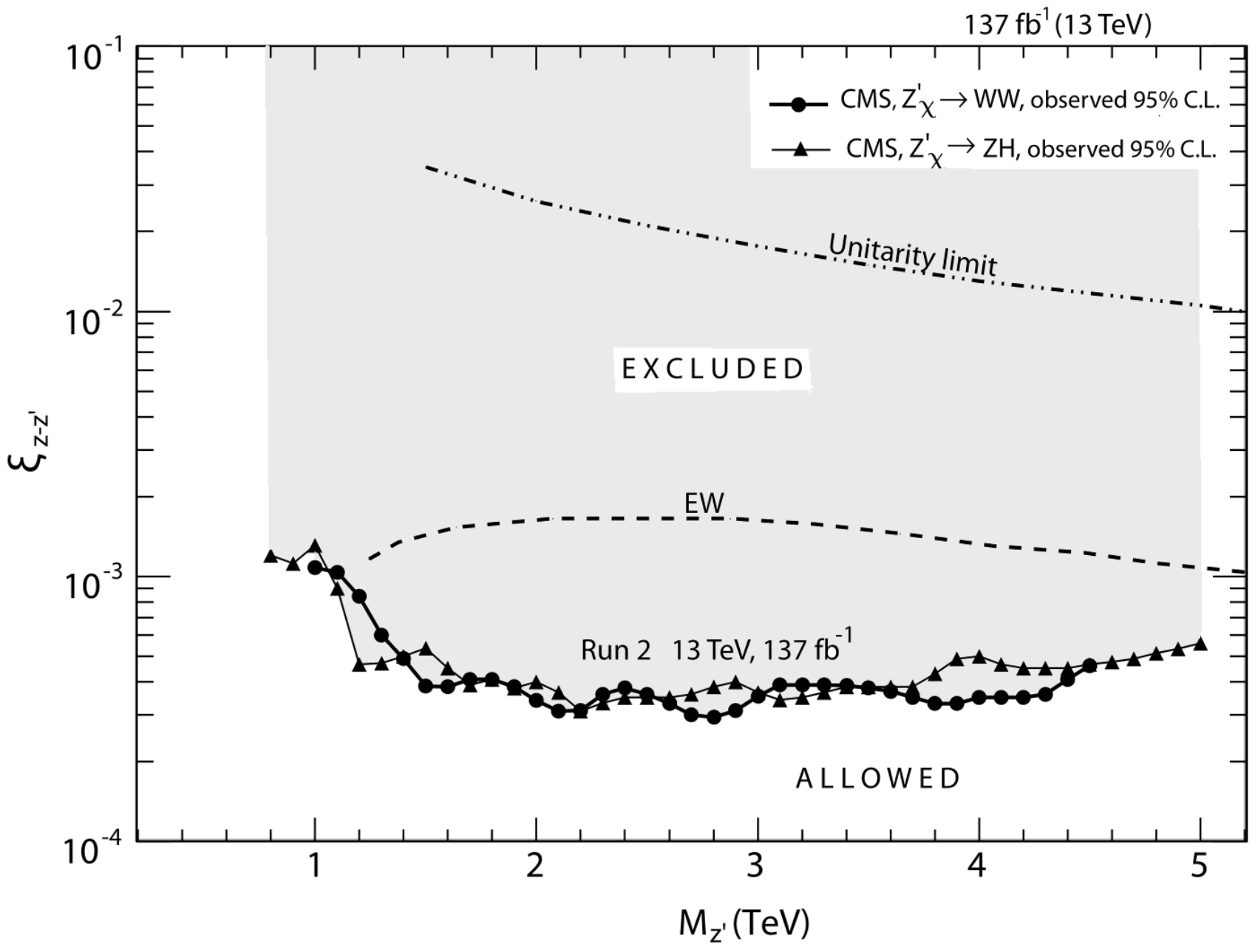}
\end{center}
\caption{
$E_6$ $Z'_{\chi}$ model: Analogous to  Fig.~\ref{Fig:cross_sect_EGM}
and Fig.~\ref{Fig:bounds-egm}, respectively.
}
\label{Fig:chi}
\end{figure}

\begin{figure}[htb]
\begin{center}
\includegraphics[scale=0.52]{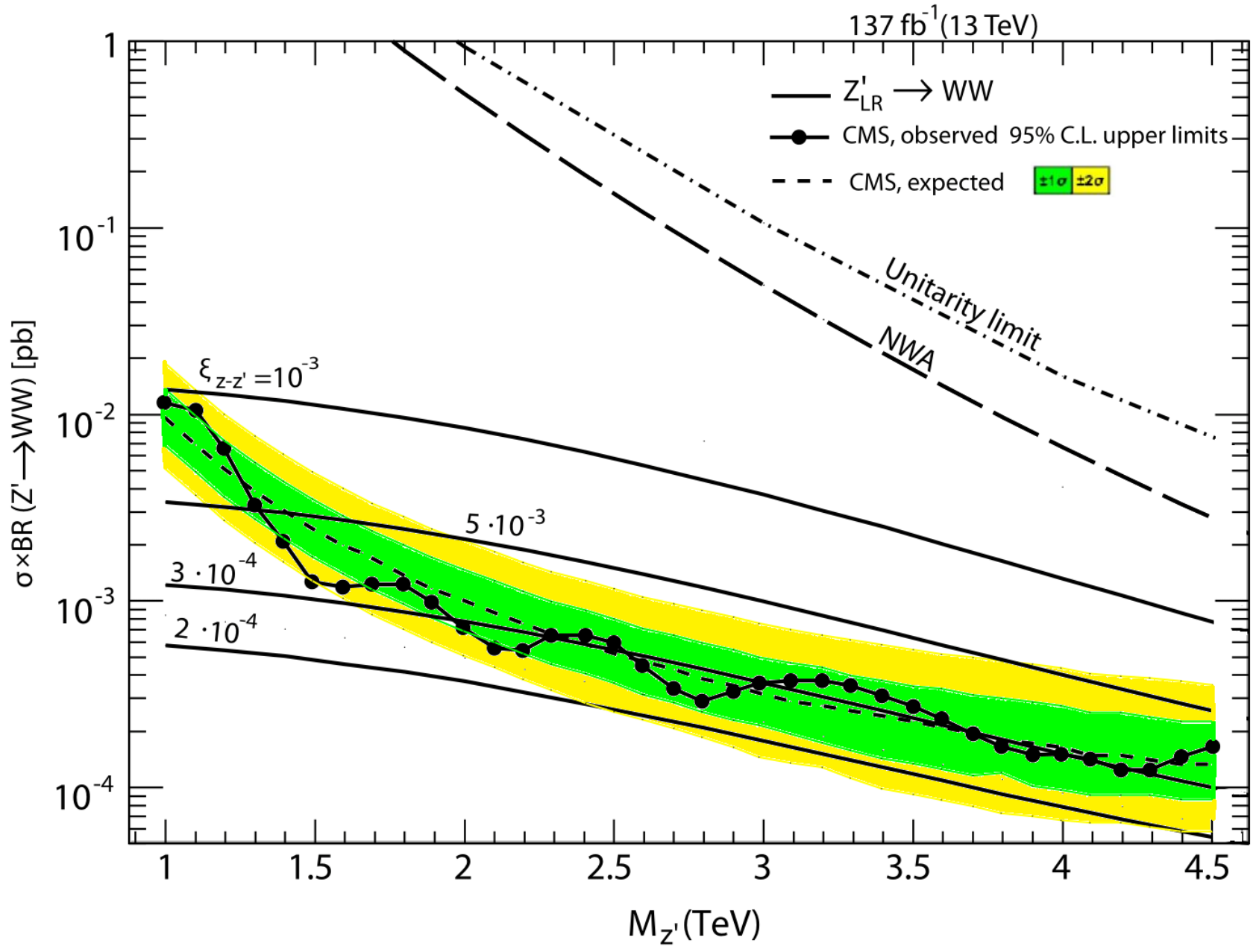}
\includegraphics[scale=0.52]{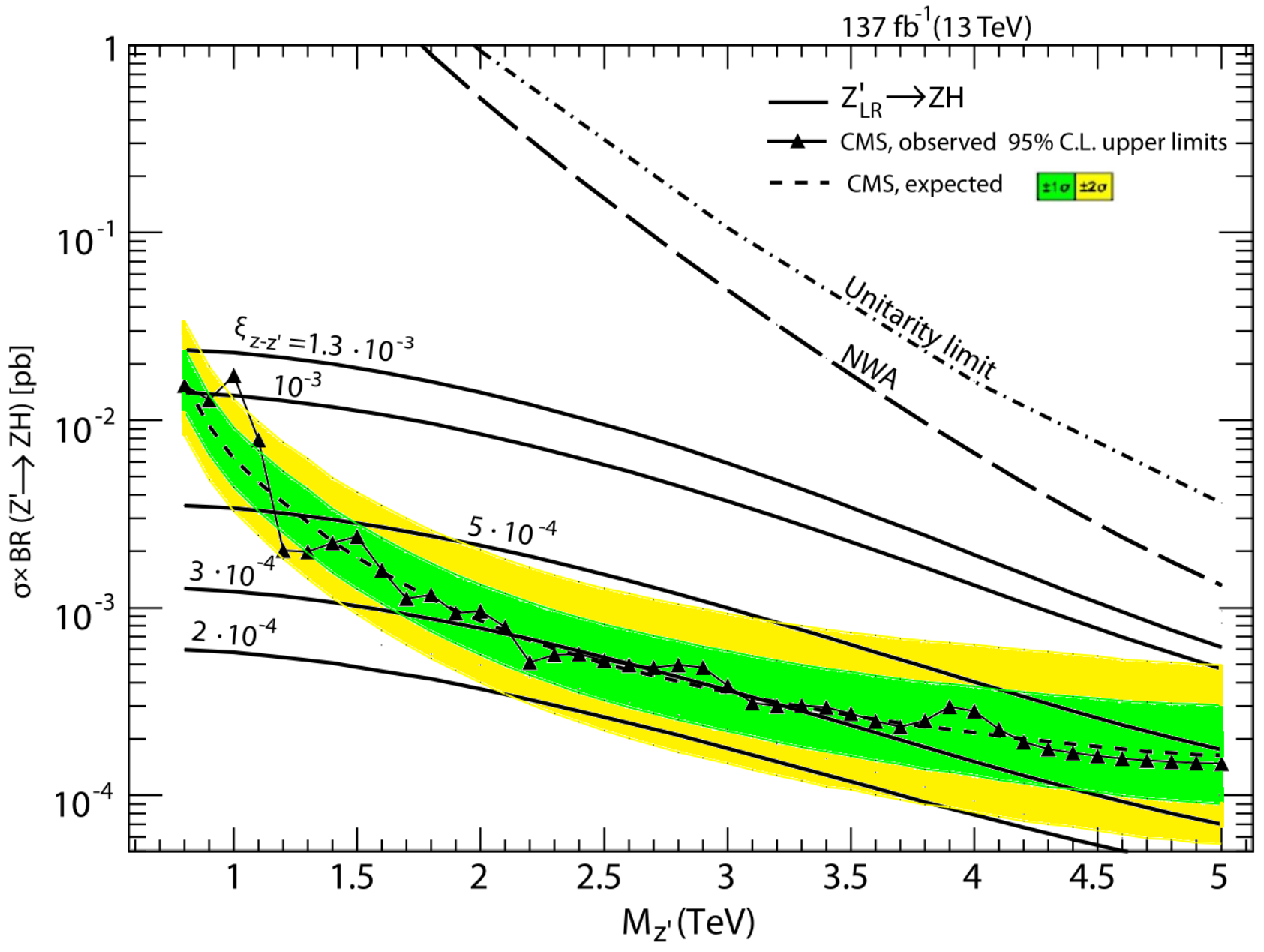}
\includegraphics[scale=0.52]{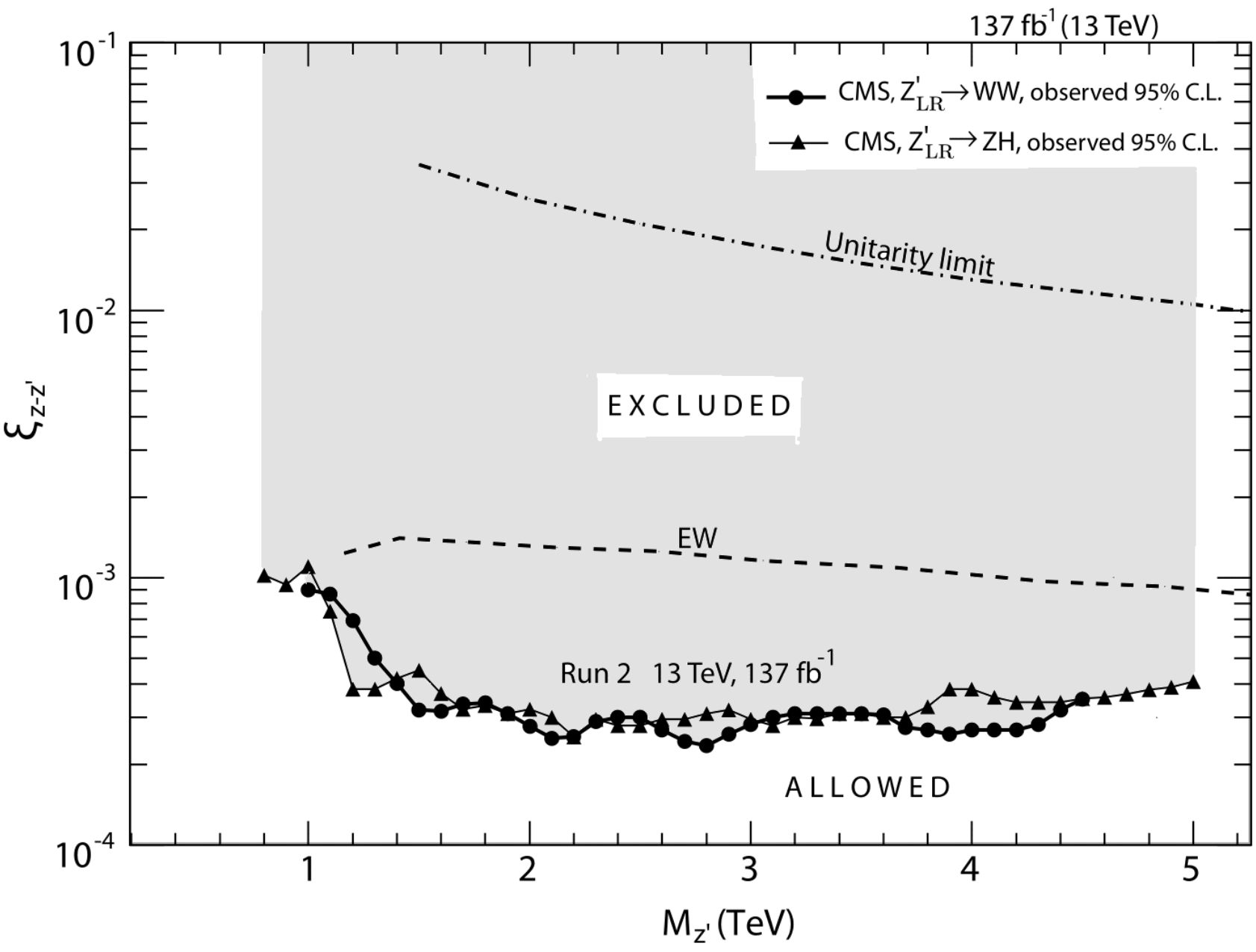}
\end{center}
\caption{
$Z'_{\rm LR}$ model: Analogous to  Fig.~\ref{Fig:cross_sect_EGM}
and Fig.~\ref{Fig:bounds-egm}, respectively.
}
\label{Fig:LR}
\end{figure}

The analysis of $Z$-$Z'$ mixing, performed here for the EGM, can also be
carried out for the other benchmark models. The results of the numerical analysis for these models are  shown in Figs.~\ref{Fig:psi}--\ref{Fig:LR}. Limits on a $Z'_{\text{model}}$ are calculated as the intersection between the observed limits, $\sigma_{95\%} \times {\rm BR}(Z'\to W^+W^-)$ and $\sigma_{95\%} \times {\rm BR}(Z'\to ZH)$,  with the model prediction, $\sigma(pp\to Z'_{\text{model}}X)\times \text{BR}(Z'_{\text model}\to W^+W^-)$ and $\sigma(pp\to Z'_{\text{model}}X)\times \text{BR}(Z'_{\text{model}}\to ZH)$, respectively.

\section{Concluding remarks}
\label{sect:conclusions}
Exploration of the diboson $WZ$/$WH$ and $WW$/$ZH$  production at the LHC with the 13~TeV
data set allows to place stringent constraints on the $W$-$W'$  and $Z$-$Z'$ mixing parameters in the resonance mass range between $\sim 1.0$~TeV and 4.5 TeV\footnote{A slightly wider resonance mass range  was taken for the process of $Z'\to ZH$, namely 0.8 -- 5 TeV.}. We derived such limits by using the full CMS Run~2 data set recorded at the CERN LHC, with integrated luminosity of 137~fb$^{-1}$.
By comparing the experimental limits to the theoretical predictions for the total cross section of the $W'$ and $Z'$ resonant production and its subsequent decay into $WZ$/$WH$ and $WW$/$ZH$ pairs,
$\sigma(pp\to W' X) \times {\rm BR}(W' \to WZ/WH)$  vs $\sigma_{95\%} \times {\rm BR}(W'\to WZ/WH)$ and $\sigma(pp\to Z' X) \times {\rm BR}(Z' \to W^+W^-/ZH)$  vs $\sigma_{95\%} \times {\rm BR}(Z'\to W^+ W^-)$, we show that the  derived limits on the mixing parameters, $\xi_{W\text{-}W^\prime}$ and $\xi_{Z\text{-}Z^\prime}$,  for the benchmark  models, are substantially improved (of the order of a few $\times 10^{-4}$) with respect to those obtained from the global analysis of low-energy electroweak data (EW), as well as from the diboson production study
performed at the Tevatron and those based on the CMS Run~1 and  on the CMS Run~2 at time-integrated luminosity of $\sim 36$ fb$^{-1}$. Further constraining of this mixing can be achieved from the analysis of future CMS data to be collected in Run~3 as well as at the next options of hadron colliders such as HL-LHC and HE-LHC  as demonstrated in \cite{Osland:2020onj} for the ATLAS experiment.

\vspace*{0mm}

\section*{Acknowledgements}

We would like to thank V.A. Bednyakov for helpful discussions. 
This research has been partially supported by the Abdus Salam ICTP (TRIL Program)
and by the Belarusian Republican Foundation for
Fundamental Research, F21ICR-001.
The work of P.O. has been supported by the Research
Council of Norway.

\bibliography{ref}
\end{document}